# Synergistic doping of the grain interior and grain boundary alters deformation mechanisms and enables extreme strength in nanocrystalline Ni-Cr-Y alloys

Yi Liu [1], Jason R. Trelewicz [2,3,4], Timothy J. Rupert [1,5,6,*]

[1] Department of Materials Science and Engineering, University of California, Irvine, CA 92697, USA

[2] Department of Materials Science and Chemical Engineering, Stony Brook University, Stony Brook, NY 11794, USA

[3] Institute for Advanced Computational Science, Stony Brook University, Stony Brook, NY 11794, USA

[4] Materials Science and Technology Division, Oak Ridge National Laboratory, Oak Ridge, TN 37831, USA

[5] Hopkins Extreme Materials Institute, Johns Hopkins University, Baltimore, MD 21218, USA

[6] Department of Materials Science and Engineering, Johns Hopkins University, Baltimore, MD 21218, USA

* Corresponding author: tim.rupert@jhu.edu

## Abstract

Solid solution addition and grain boundary segregation have been independently shown to enhance the strength of nanocrystalline alloys. In the present study, the interplay between these two effects is investigated in nanocrystalline Ni-Cr-Y sputtered films through systematic variation of alloying element contents with grain size kept constant. Cr is introduced into a solid solution and serves to strengthen the lattice, while Y segregates to the grain boundaries to stabilize these features. Nanoindentation is used to probe hardness, with unexpected trends and very high values observed. Cr additions led to nanocrystalline solid solution strengthening, yet saturation was observed at higher concentrations due to the emergence of grain boundary dominated processes, as evidenced by pile-up morphologies containing slip steps and grain rotation. Y enrichment was observed at the examined grain boundaries, enhancing boundary-mediated strengthening by pinning the dislocations and suppressing dislocation emission, grain boundary sliding, and grain rotation processes. With increasing Y concentration, the nanocrystalline solid solution strengthening effect induced by Cr addition becomes weaker. This phenomenon can be attributed to a reduced dislocation bowing distance caused by dopant pinning. Most notably, the strongest ternary Ni-Cr-Y alloy exhibited a hardness of 11.0 GPa, among the highest hardness values reported for single-phase nanocrystalline Ni alloys with grain sizes near 20 nm. These findings highlight how tuning grain and grain boundary chemistry offers a viable strategy to control dislocation mechanics and improve the strength of nanocrystalline metals.

## 1 Introduction

Achieving high strength in metallic materials remains a critical objective in alloy design, particularly for applications where mechanical reliability under load is paramount. Nanocrystalline metals have garnered significant research interest due to their exceptionally high strength and distinctive deformation mechanisms [1–4]. As grain sizes are refined to the nanometer regime, the classical Hall-Petch grain-size strengthening effect breaks down [5, 6]. The Hall-Petch breakdown occurs as conventional dislocation-mediated plasticity becomes increasingly suppressed, giving way to grain-boundary-dominated mechanisms such as grain boundary sliding [7], grain boundary migration [8], grain rotation [9], and grain boundary dislocation emission [10–12]. While these nanoscale processes enable high strength, the achieved strengths remain far below the theoretical limit for metallic crystals, indicating substantial room for further improvement. At the same time, these deformation pathways can promote strain softening and mechanical instability, especially in alloys with extremely small grains. Consequently, identifying strengthening strategies that explicitly account for grain boundary-centric deformation mechanisms is essential for fully exploiting the advantages of nanocrystalline structures.

During the plastic deformation of nanocrystalline metals, dislocations no longer operate as traditional sources within grain interiors but instead nucleate, emit, and are absorbed at grain boundaries [12–15]. Furthermore, grain boundary processes such as sliding and rotation can become active [16, 17]. With these new mechanisms, the local chemistry and structure of grain boundaries can strongly influence the activation of the grain-boundary-dominated deformation mechanisms. Dopant atoms that segregate to grain boundaries may act to alter local stress states, increasing the barrier for dislocation processes such as nucleation, emission and propagation [18, 19]. Grain boundary processes, including sliding and rotation, can also be affected by chemical and structural modification of the boundary region [20–25]. For example, Qian et al. [20] reported that Zr segregation in nanocrystalline Cu with grain sizes below 5 nm led to a strength of ~2.6 GPa, significantly higher than the ~1.85 GPa of undoped Cu with similar grain size. This enhancement was attributed to Zr-induced amorphization of the grain boundary that suppressed dislocation-grain boundary interactions and hindered plasticity. Similarly, Masuda et al. [26] demonstrated that grain boundary segregation of Cu and Mg in ultrafine-grained A2024 alloys can contribute up to ~565 MPa to the yield strength. Cu and Mg segregation to the grain boundaries was found to effectively impede dislocation motion and stabilize the ultrafine-grained structure by relaxing the local strain around dislocations. In these cases, the dislocation and grain boundary mechanisms during deformation were restricted so that the overall strength and hardness were increased.

Solid solution strengthening has also been shown to be active in nanocrystalline alloys [3, 27, 28] but with a number of interesting behaviors emerging that generally are not observed in coarse-grained alloys. Randomly distributed solute atoms interact with freely migrating dislocations in coarse-grained materials, with elastic misfit and modulus differences being of particular importance. Among several classical models, the Fleischer model [29] provides a phenomenological description of solid solution strengthening where the strength of the alloy scales with the square root of solute concentration. For nanocrystalline systems, the effect of solutes is fundamentally altered and new opportunities arise [3]. Due to the limited migration distances and the high-volume fraction of grain boundaries, dislocations become heavily pinned by grain boundaries and the critical dislocation bowing distance is often the grain size. Rupert et al. [3] developed a nanocrystalline solid solution strengthening model based on Ni-W alloys that had two components: (1) a nanocrystalline solution effect from changes to the overall grain stiffness and lattice constant and (2) an individual solute atom pinning effect from traditional contributions on the local scale. These authors found that nanocrystalline solid solution strengthening could provide more pronounced strengthening effects in some cases, yet that solid solution softening was also possible. Kim et al. [30] investigated solid solution strengthening in nanocrystalline Ni alloyed with Mo and W. These authors found that the hardness increased nearly linearly with W content, from 6.85 GPa in pure nanocrystalline Ni with an average grain size of 15 nm [31] to almost 11 GPa at Ni-25Mo-5W (at.%) with an average grain size of a few nanometers. Such strengthening is much greater than predicted by the classical Fleischer model, underscoring that nanocrystalline solid solution strengthening offers new opportunities for improving mechanical properties.

Based on the discussion above, (1) grain boundary segregation and (2) nanocrystalline solid solution additions can be thought of as two mechanistically distinct pathways for enhancing the strength of nanocrystalline metals. In this study, we investigate a ternary nanocrystalline Ni-Cr-Y alloy system to systematically examine the interplay between the two strategies. The concentrations of Cr and Y were independently varied, with Cr serving as a lattice-soluble solute to strengthen grain interiors by increasing the stiffness and Y selected as a grain boundary segregant. The alloys were deposited by magnetron sputtering, yielding single-phase face-centered cubic (FCC) structures with consistent average grain sizes of 20-30 nm. Nanoindentation measurements revealed hardness values up to 11.0 GPa, among the highest reported for Ni-based nanocrystalline alloys in general and the highest value for this grain size. Both Cr and Y increased hardness, but their effects were not independent. Y segregation raised the initial and high-Cr hardness levels and widened the Cr composition window for effective solid solution strengthening, while decreasing the Cr-hardening slope. The reduced slope indicates a competitive effect on the marginal contribution of Cr, whereas the increased hardness levels and extended Cr-strengthening range represent a cooperative system-level contribution. Here, "synergistic" refers to overall strengthening effects that are larger than existing theoretical predictions due to the emphasis on dislocation emission. A plateau in strength is found in some cases, with post mortem investigation suggesting that this behavior is related to a shift from dislocation-based mechanisms to collective grain boundary plasticity. In general, this work demonstrates that coupling of nanocrystalline solid solution strengthening with grain boundary segregation strengthening provides a viable route for accessing extreme strength and altering the dominant deformation mechanisms of nanocrystalline metals.

## 2 Materials and methods

### 2.1 Materials selection

Alloying elements were carefully selected to activate each mechanism independently. For solid solution strengthening, alloying elements were chosen based on the modified nanocrystalline solid solution strengthening model of Rupert et al. [3], which incorporates Burgers vector and modulus mismatches between solute and solvent atoms. To ensure a positive strengthening contribution, the solute should possess a larger shear modulus and a larger atomic size relative to the base metal. A sufficiently high solubility limit is also required to maintain a stable single-phase solid solution, which was assessed using phase diagrams and thermodynamic criteria [32]. Based on these considerations, Ni was chosen as the matrix and Cr as the primary solute due to its high solubility in Ni and its favorable shear modulus and lattice mismatches that promise strengthening [33–36]. Cr concentrations from 0 to 24 at.% were investigated. In contrast, grain boundary segregation strengthening requires a solute with limited bulk solubility but strong interfacial segregation tendency. For transition metals, this typically corresponds to elements with large elastic enthalpies to give a strong thermodynamic driving force for lowering grain boundary energy upon segregation [32]. Y was selected to fulfill this role, as it exhibits a substantial size mismatch with Ni, limited lattice solubility, and has been previously reported to segregate to grain boundaries in similar alloy systems [37]. Overall Y concentrations of 0.5, 1.5, and 3 at.% were studied. As the grain boundary region can become an appreciable portion of the overall volume fraction for very fine grain sizes (see, e.g., [38]), grain boundary concentrations are expected to be much higher than the global concentration.

## 2.2 Nanocrystalline solid solution alloy fabrication

Nanocrystalline Ni-Cr-Y thin films were deposited by magnetron co-sputtering onto single crystal quartz substrates with a (0001) orientation (UniversityWafer, USA) using an AJA Orion-8 sputtering system (AJA International, USA). High-purity elemental targets were employed for Ni (99.999%, AJA International, USA), Cr (99.95%, Kurt J. Lesker Company, USA), and Y (99.9%, Kurt J. Lesker Company, USA). The chamber was pumped to a base pressure below $5 \times 10^{-7}$ Torr and then a constant argon gas flow of 50 standard cubic centimeters per minute (sccm) was introduced to maintain a working pressure of 4 mTorr during deposition. Substrate rotation was employed throughout to ensure uniformity in film thickness and composition. Deposition rates for each target were first calibrated using a quartz crystal thickness monitor. Subsequently, the sputtering power for each gun was adjusted to achieve the desired alloy compositions. Final film thicknesses were in the range of 600-650 nm. It has been reported that inverse Hall-Petch behavior occurs as the grain size of Ni drops below ~10 nm [6, 39]. Thus, a relatively constant grain size above this level is targeted here and achieved with substrate heating during deposition and post-deposition annealing treatments, when needed. Samples containing 0.5 and 1.5 at.% Y were deposited at 400 °C, which yielded stable grain sizes in the target range. For the 3 at.% Y alloy, films were deposited at 600 °C followed by an 8 h anneal at the same temperature. The higher temperatures were needed due to the strong grain boundary segregation tendency of Y, which suppresses grain growth [40]. Across all samples, grain sizes were maintained in a range of 20-30 nm, with details provided below.

### 2.3 Microstructural characterization

The phase content of the films was characterized using grazing incidence X-ray diffraction (XRD) at the grazing incident angle of 1° on a Rigaku SmartLab X-ray diffractometer, with a Cu Kα radiation source operated at 40 kV and 44 mA and a point scintillator counter. 2θ scans were collected over a range of 30°-80° with a step size of 0.01° and a scan speed of 1°/min. Lattice parameters were calculated from the peak position of the strongest diffraction peak using Bragg's law. Peak broadening was analyzed using the Scherrer equation, with a shape factor of 0.9 and instrumental broadening subtracted using a $LaB_6$ standard, to estimate volume-averaged grain size ($d$). Film surface morphology and chemical composition were examined using a field-emission scanning electron microscope (SEM) (FEI Magellan 400 XHR) equipped with an Oxford 80 $mm^2$ silicon drift energy dispersive X-ray spectroscopy (EDS) detector. Additional nanoscale surface topography measurements were performed using an Anton Paar Tosca 400 atomic force microscope (AFM) operated in tapping mode.

Transmission electron microscopy (TEM) was carried out using a JEOL JEM-2800 operated at 200 kV and equipped with a Gatan OneView camera. Bright-field TEM (BF-TEM) imaging was used to assess grain shapes and sizes. Grain sizes were determined by manually outlining more than 50 resolvable contrast-defined crystalline domains and calculating their circular equivalent diameters. Clearly resolved twins were included when they could be distinguished. High-resolution characterization of grain boundary structure and chemistry was performed using a JEOL JEM-ARM300F Grand ARM equipped with double spherical aberration (Cs) correctors and dual 100 $mm^2$ silicon drift EDS detectors, operated at 300 kV. Atomic-resolution high-angle annular dark-field scanning TEM (HAADF-STEM) imaging was conducted with a probe current of 35 pA, using inner and outer collection angles of 64 and 180 mrad, respectively. Atomic-resolution bright-field STEM (BF-STEM) imaging was conducted with the same probe current and outer collection angles of 38 mrad. Grain boundary chemistry was characterized by using high resolution EDS measurement with a probe current of 204 pA. All TEM lamellae were prepared and thinned to electron transparent by focused ion beam (FIB) lift-out method [41] using a FEI Quanta 3D FEG dual-beam SEM/FIB microscope with an OmniProbe micromanipulator. Final polishing was performed at 5 kV and 48 pA to minimize beam damage.

### 2.4 Mechanical testing

Nanoindentation hardness measurements were performed using a KLA G200 Nanoindenter equipped with an Enhanced Dynamic Contact Module II (DCM II) and a Berkovich diamond tip. Indentation was carried out in quasi-static mode under depth control, with the maximum indentation depth fixed at 100 nm. This depth was selected with reference to established relationships between the maximum penetration depth and plastic-zone size in nanoindentation [42]. Because 100 nm corresponds to approximately 15-17% of the film thickness and analytical estimates alone cannot exclude substrate effects, possible substrate interference was further evaluated experimentally using depth-dependent hardness measurements, as described below. A constant loading rate of 0.075 mN/s was used. At the peak load, the indenter was unloaded to 80% of the maximum value and held briefly to monitor thermal drift (data were excluded if the drift rate exceeded 0.1 nm/s), before completing the unloading segment. For each composition, at least 100 indents were performed with a minimum distance between neighboring indents of 5 µm, approximately 50 times the indentation depth, to avoid overlapping deformation fields. Hardness values were extracted from load-displacement curves using the Oliver-Pharr method [43], with the tip area function calibrated using a fused silica standard prior to testing. Only tests containing a clear measurement artifact were excluded, whereas no measurement was removed solely on the basis of its hardness value or an IQR-based statistical criterion. The individual hardness measurements and their statistical distributions, including the mean, standard deviation, median, and interquartile range, are presented in Figure S1. All tests were performed at room temperature. Possible substrate interference was evaluated using a separately fabricated Ni-2.0Y film with the same film thickness, quartz substrate, deposition procedure, and indentation protocol. Hardness measurements were performed at maximum indentation depths of 50, 80, 100, 150, 200, 300, and 400 nm. As shown in Figure S2, the measured hardness remained constant within experimental scatter between 50 and 100 nm and then decreased at greater depths. The maximum depth of 100 nm used for the compositional study was therefore selected to probe the most material possible while avoiding substrate-induced softening. Representative load-displacement curves for low-, intermediate-, and high-hardness alloys are provided in Figure S3.

In addition to Berkovich indentation, selected samples were tested using a cube corner diamond tip to assess their resistance to localized plastic deformation. The cube corner tests were performed under the same depth-controlled quasi-static conditions, with a maximum indentation depth of 600 nm and the same loading rate. Due to the sharper geometry of the cube corner tip, these tests generated higher stress concentrations and were used to qualitatively evaluate the plastic deformation behavior of the films under extreme loading conditions. No mechanical properties were extracted from the experiments with a cube corner tip.

## 3 Results and Discussion

This section presents a systematic investigation of the microstructure and mechanical behavior of nanocrystalline Ni-Cr-Y thin films across a range of dopant concentrations. Section 3.1 establishes the crystal structure, elemental homogeneity, and grain size of the deposited films using XRD, SEM-EDS, and TEM. Section 3.2 examines grain boundary structure and chemistry at atomic resolution with TEM, revealing Y segregation to the boundary region. Section 3.3 presents nanoindentation hardness data as a function of composition and interprets the observed strengthening trends through nanocrystalline solid solution strengthening models. Section 3.4 discusses the hardness saturation behavior and connects it to a transition in the dominant deformation mechanism through post-indentation morphological and microstructural analysis. Throughout, representative samples from the compositional series are shown in each subsection to provide illustrative examples of the broader microstructural and mechanical trends, with the complete hardness dataset spanning all compositions shown in Section 3.3.

### 3.1 Microstructure of Ni-Cr-Y thin films

Figures 1(a), (b) and (c) show a representative surface from a Ni-9.3Cr-1.5Y (at.%) alloy sample, chosen to show here because it has an intermediate amount of both dopant species. The film surface is dense and free of pores, cracks, or delamination. In Figure 1(c), the high magnification SEM image reveals that the surface consists of well-defined grain-like clusters separated by distinct boundaries. These features can be interpreted as the top-view expression of columnar grains commonly formed during magnetron sputtering [44, 45]. This uniform and compact morphology indicates that the deposited films are suitable for subsequent nanoindentation measurements to evaluate mechanical properties. Figures 1(d)-(f) show EDS elemental maps of Ni, Cr, and Y, respectively, for the same sample. The maps reveal homogeneous distributions of all three elements across the film surface, with no evidence of secondary phase formation within the spatial resolution of the SEM-EDS system. Quantitative EDS results for all samples are summarized in Table 1.

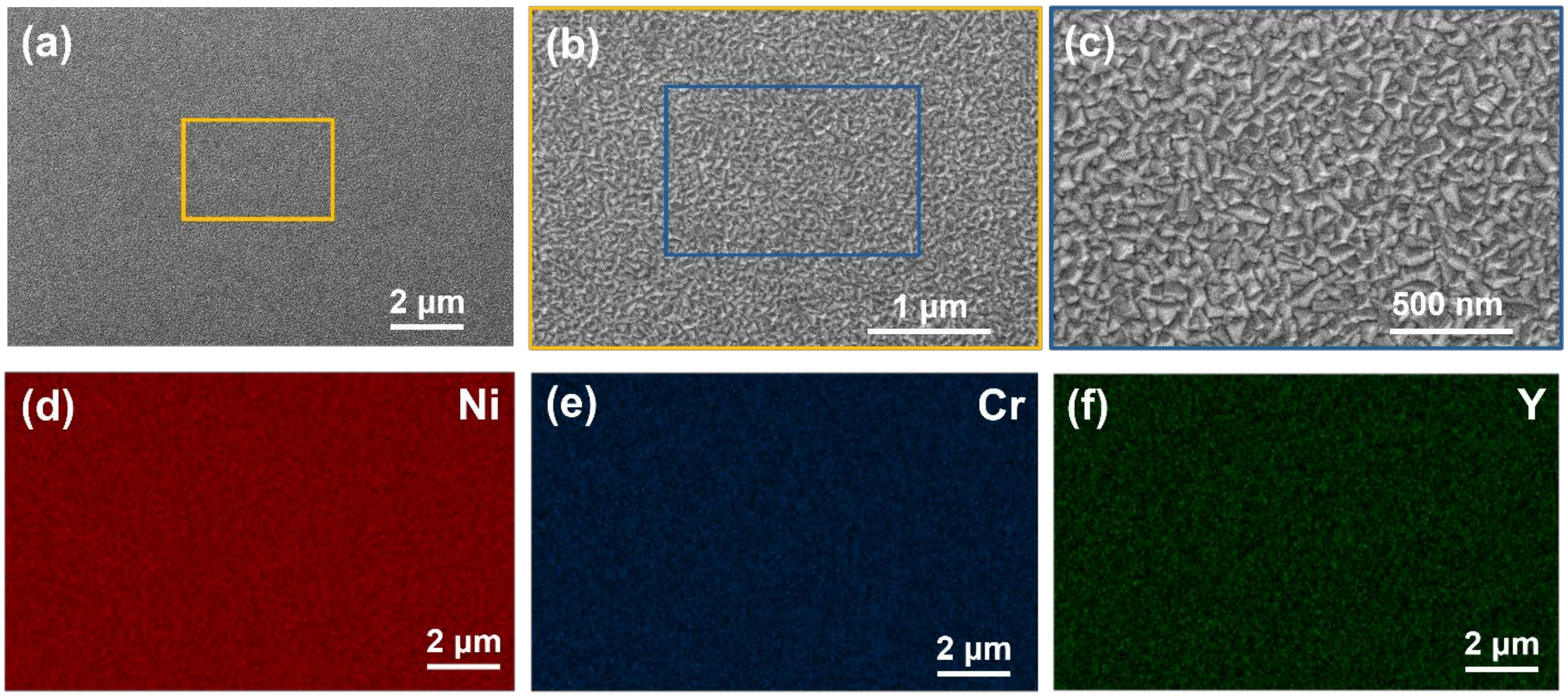

**Figure 1** Surface morphology and elemental distribution of the Ni-9.3Cr-1.5Y (at.%) sample. **a-c** SEM images acquired at low and high magnifications, show a dense surface and cluster features. EDS elemental maps of **d** Ni, **e** Cr, and **f** Y confirm uniform macroscopic distribution without evidence of phase separation.

| EDS Composition (at.%) | | | XRD Grain Size (nm) | XRD Lattice Parameter (Å) |
|---|---|---|---|---|
| Ni | Cr | Y | | |
| 99.5 | 0 | 0.5 | 26 | 3.522 |
| 98.3 | 1.2 | 0.5 | 25 | 3.524 |
| 94.9 | 4.6 | 0.5 | 27 | 3.527 |
| 89.2 | 10.3 | 0.5 | 30 | 3.533 |
| 85.8 | 13.7 | 0.5 | 26 | 3.536 |
| 98.5 | 0 | 1.5 | 26 | 3.525 |
| 92.8 | 5.9 | 1.3 | 25 | 3.528 |
| 89.2 | 9.3 | 1.5 | 25 | 3.531 |
| 84.0 | 14.5 | 1.5 | 25 | 3.537 |
| 74.9 | 23.6 | 1.5 | 27 | 3.550 |
| 97.0 | 0 | 3.0 | 20 | 3.526 |
| 88.7 | 8.3 | 3.0 | 20 | 3.530 |
| 81.0 | 16.0 | 3.0 | 21 | 3.546 |
| 74.7 | 22.5 | 2.8 | 20 | 3.550 |

**Table 1** EDS-measured concentrations, XRD-derived grain sizes, and XRD-calculated lattice parameters for all Ni-Cr-Y samples investigated in this study. Grain sizes remain within the nanocrystalline regime (20-30 nm), enabling meaningful comparison of strengthening mechanisms.

A set of representative XRD patterns are shown in Figure 2(a), where all samples only exhibit reflections corresponding to the FCC Ni-rich phase. The (111), (200), and (220) peaks are present and remain symmetric throughout the series, with no detectable secondary phases, intermetallic compounds, or amorphous features. The absence of peak splitting or new reflections indicates that no phase separation or ordering occurs within the resolution of the measurement, and the FCC structure is retained upon both Cr and Y incorporation. We note that a slight asymmetry in the (200) reflection is observed, which is the second peak in Figure 2(a). This has previously been shown to result from planar defects such as stacking faults or twin boundaries [46], a topic that will be discussed further in the next section.

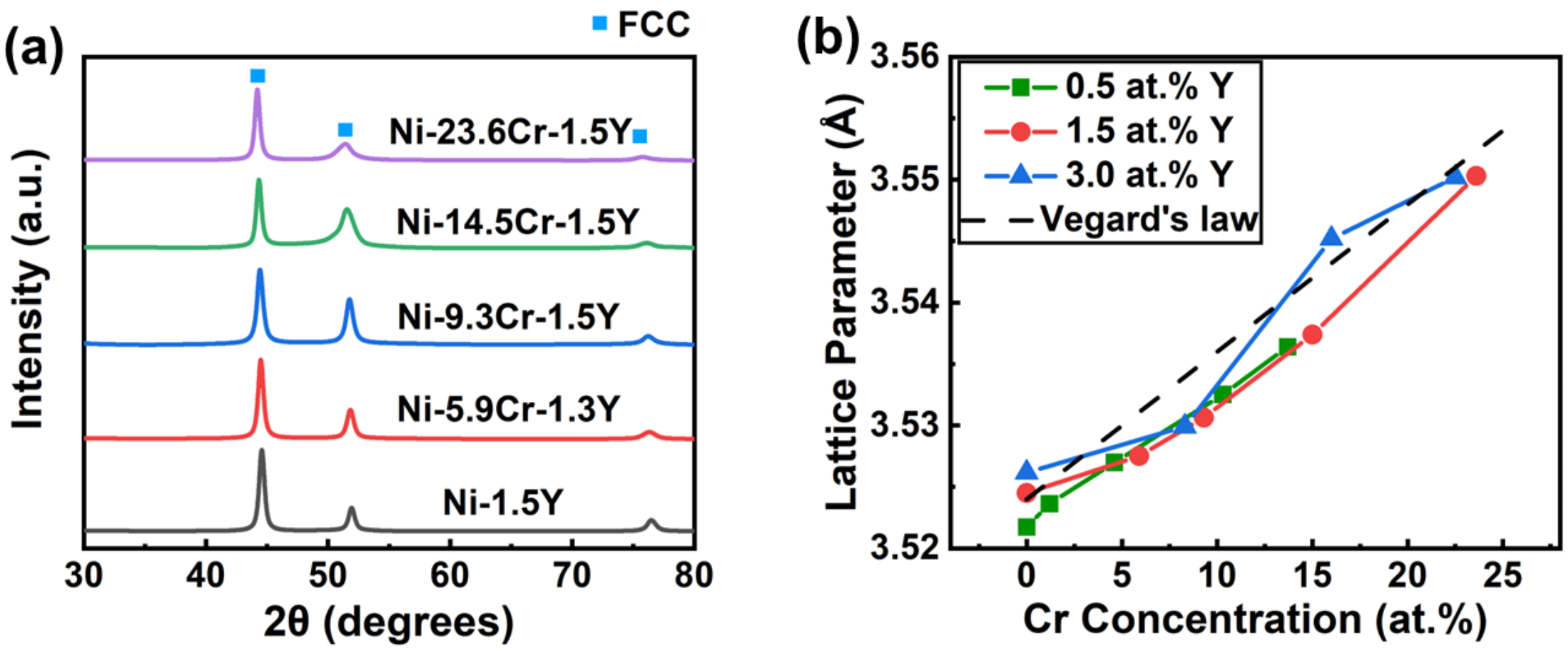

**Figure 2** **a** Representative grazing incidence XRD patterns from the Ni-Cr-Y samples with 1.5 at.% Y, showing a single-phase FCC solid solution without secondary phases. **b** Lattice parameters extracted from the XRD data as a function of Cr concentration, which generally match Vegard's law [47], indicating lattice expansion due to Cr incorporation into the Ni lattice.

The effect of Cr and Y additions on the lattice structure is further analyzed through the evolution of lattice parameters, shown in Figure 2(b). With increasing Cr content, the diffraction peaks continuously shifted toward lower 2θ values, corresponding to an increase in lattice parameter (also included in Table 1), which indicates lattice expansion due to Cr addition. The values extracted from XRD are in close agreement with the theoretically estimated lattice parameters based on Vegard's law [47]. In contrast, the datasets for samples with different Y content all fall on top of one another, showing that addition of Y has no influence on the lattice parameter despite its large atomic size. Because the lattice parameter is obtained from the coherent FCC Bragg peaks, any Y dissolved in the grain interiors would shift these peaks; the absence of such a shift indicates that Y may reside preferentially at the grain boundaries. Direct evidence of this grain boundary segregation is provided below with STEM-EDS.

Residual stresses were not directly measured in this study. However, deposition temperatures of 400 and 600 °C correspond to homologous temperatures of $T/T_m$=0.39 and 0.51 for Ni, respectively, where enhanced adatom and grain boundary diffusion are expected to relax a substantial fraction of the intrinsic sputtering stress. The additional 8 h anneal applied to the 3 at.% Y films would provide further opportunity for stress relaxation, consistent with previous observations for sputtered Ni films after high-temperature annealing [48]. In addition, the Cr-dependent hardening trends are established within individual Y series with identical thermal histories, so residual stress cannot be the explanation for any strength trends.

Grain size estimates from XRD were obtained using the Scherrer equation applied to the strongest (111) peaks, with instrumental broadening corrected using a $LaB_6$ standard. The resulting volume-averaged grain sizes were found to lie in the range of 20-30 nm across all samples, as summarized in Table 1. The complete grazing-incidence XRD patterns are provided in Figure S4. All films retain the same single-phase FCC Ni-rich structure without detectable secondary-phase reflections, and no systematic Y-series-dependent change in the relative intensities of the (111), (200), and (220) reflections is observed. The slightly broader reflections of the 3 at.% Y series are consistent with its marginally smaller crystallite sizes of 20-21 nm, compared with 25-30 nm for the other series. Together with the common Cr-dependent Vegard trend in Figure 2(b), these observations indicate broadly comparable phase constitution, texture, and lattice strain behavior among the three series within the sensitivity of the present measurements. Nevertheless, the different thermal history of the 3 at.% Y series could possibly influence defect density, grain boundary structure, residual stress, impurity content, and segregation state in ways that cannot be independently resolved by the present XRD measurements. Absolute property differences involving this series should therefore be interpreted as reflecting the combined effects of Y concentration and processing history. In contrast, the Cr-dependent trends within each Y series and the comparison between the identically processed 0.5 and 1.5 at.% Y series are not subject to this processing-history difference.

These grain size estimates were further validated with BF-TEM imaging. Figure 3 shows representative BF-TEM images of Ni-1.5Y and Ni-9.3Cr-1.5Y samples. The images indicate that both films exhibit columnar growth structures typical of sputter-deposited thin films [44, 49–51]. Importantly, there are many nanocrystalline grains within each column that are very fine and have equiaxed shapes. This observation confirms that the films are composed of densely packed nanocrystalline grains separated by numerous grain boundaries, thereby providing a large interfacial area for potential grain boundary segregation. Nanotwins were observed in both representative samples, with a slightly higher density in Ni-9.3Cr-1.5Y than in Ni-1.5Y. This observation is consistent with the increasingly asymmetric shoulder of the (200) XRD reflection at higher Cr concentrations. The grain size measurements in Figures 3(c) and 3(f) yielded an average circular equivalent diameter of approximately 22 nm for both samples. To further quantify the twin structure, the thickness and width of 50 resolved twins were measured in representative BF-TEM images of each sample. The mean twin thicknesses were 3.9 ± 1.7 nm for Ni-1.5Y and 3.5 ± 1.7 nm for Ni-9.3Cr-1.5Y, while the corresponding mean widths were 30.8 ± 8.7 nm and 41.3 ± 16.0 nm, respectively (Figure S5). Thus, the twin sizes are very similar, with the twins in the Cr-containing sample being slightly wider. Because the twins are much wider than they are thick, using the twin thickness alone as a grain dimension would exaggerate their effect on the grain size distribution and both dimensions must therefore be considered when accounting for the subdivision produced by twins. These results indicate that the greater apparent prevalence of nanotwinning with Cr addition is signaled by increased twin width rather than reduced twin size.

Twin boundaries may therefore provide an additional strengthening contribution, but the present results do not indicate that nanotwinning is the sole or dominant origin of the hardness trend. The measured average grain sizes from TEM were 22 nm for both samples, which are in reasonable agreement with the XRD-estimated values. This consistency suggests that the XRD-based grain size estimates are reliable. Furthermore, this grain size lies marginally above the regime where the Hall-Petch relationship typically begins to break down and grain boundary-mediated mechanisms start to strongly affect the deformation behavior [6, 15, 39].

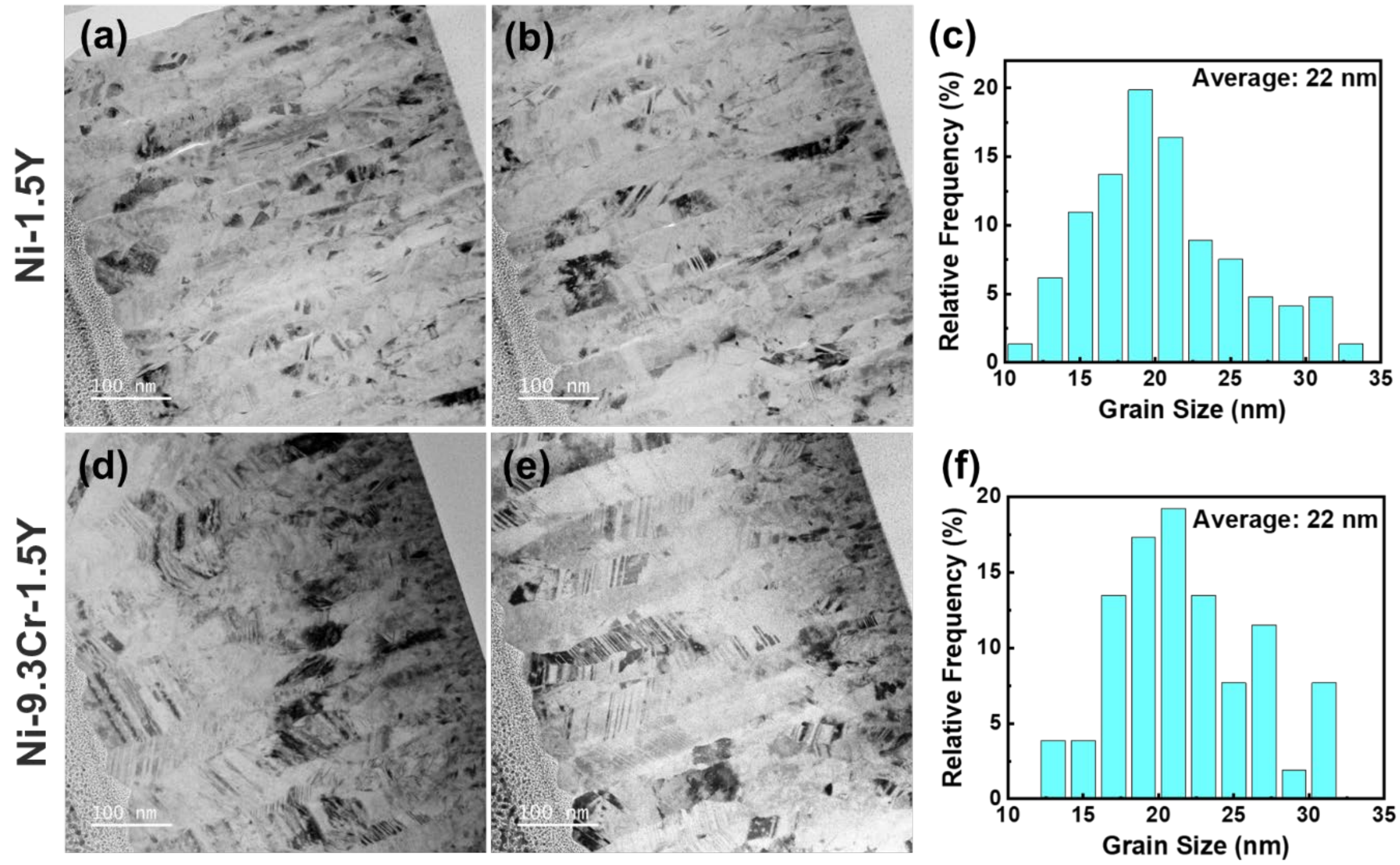

**Figure 3** Representative cross-sectional TEM images of nanocrystalline **a** and **b** Ni-1.5Y, **d** and **e** Ni-9.3Cr-1.5Y samples. **c** and **f** Grain size distributions obtained from more than 50 measurements across multiple cross-sectional regions confirm average grain sizes that are constant and similar to the XRD measurements. All data were acquired from through-thickness areas to ensure statistical representativeness of the entire film microstructure. The reported grain sizes were measured from resolvable contrast-defined regions, to account for all boundaries that can act as obstacles to dislocation motion.

### 3.2 Grain boundary structure and chemistry

Fig. 4a presents a BF-STEM image of the Ni-13.7Cr-1.5Y sample where the crystalline nature of the grains can be seen more clearly. The boundary is structurally ordered, with no indications of grain boundary films or second-phase precipitates. Fig. 4b provides a HAADF-STEM image from the same location, where a sharp contrast transition is observed near the grain boundary. Atomic columns are clearly resolved on the right side of the interface, while the left side exhibits reduced contrast due to slight misalignment from the zone axis. STEM-EDS elemental mapping and line scan results, taken from the location denoted by a yellow box in Figure 4(c) and shown in Figures 4(d)-(g), provide insight into the chemical nature of the boundary. The elemental maps and line scan profile across the interface reveal significant enrichment of Y along the grain boundary, while Ni is depleted at the grain boundary and Cr remains uniformly distributed throughout the microstructure. The line scan profile across the interface confirms that there is Y segregation, increasing from a matrix baseline of <1 at.% to a local maximum of 5.0 at.%. The absence of second-phase formation or clustering further indicates that Y remains in a dispersed, grain-boundary-segregated state. According to the Langmuir-McLean model, the presence of a strong segregation peak at the boundary implies a negative segregation enthalpy, which corresponds to a lowering of interfacial energy.

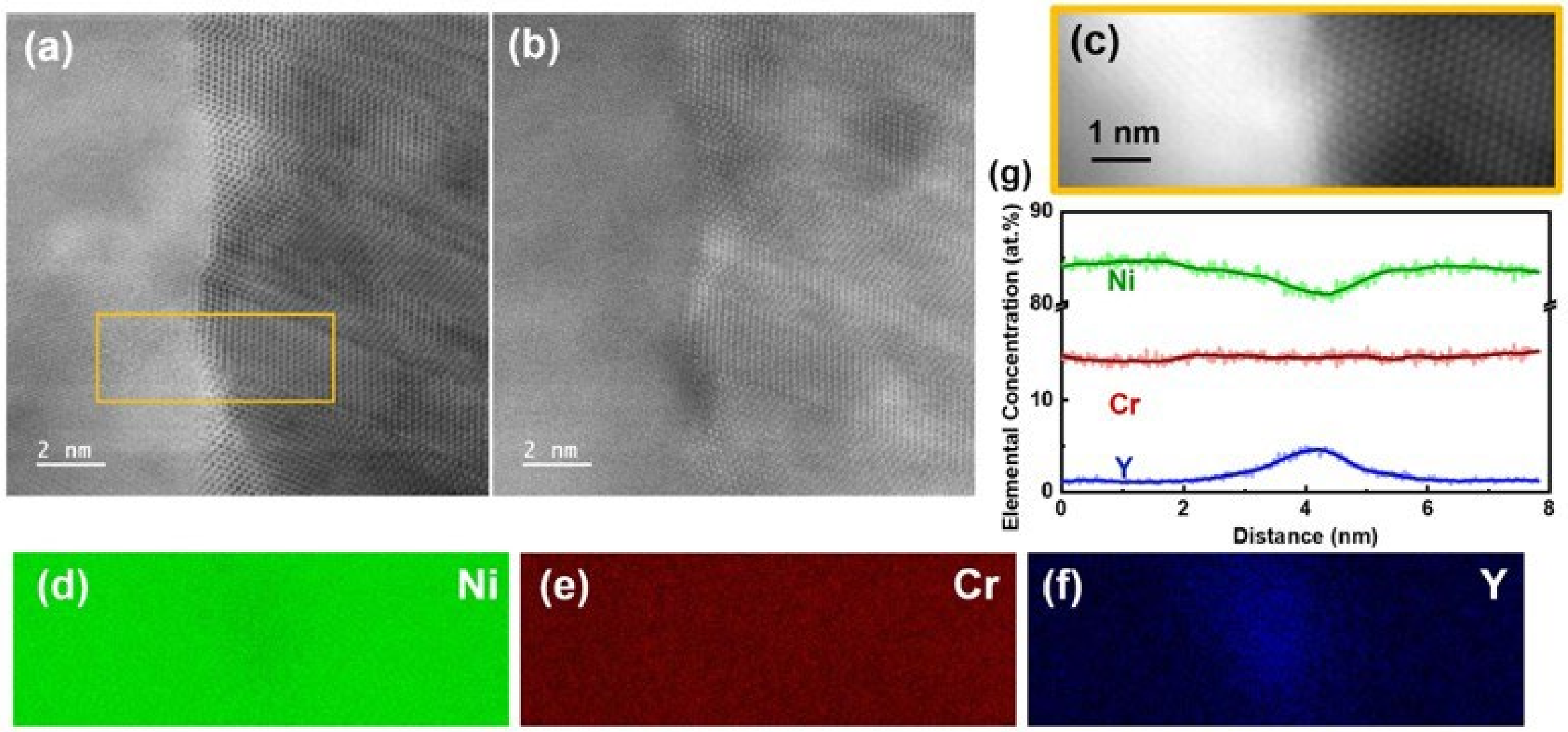

**Figure 4** Structural and compositional analysis of a grain boundary in the Ni-13.7Cr-1.5Y representative sample. **a** BF-STEM image of the grain boundary, revealing the crystalline structure at the boundary. **b** HAADF-STEM image acquired from the same location as **a**, showing a contrast transition between a zone-axis-aligned grain (right) and a slightly misaligned grain (left). **c** Magnified view of the boxed region in **a**, highlighting the atomic structure at the boundary. **d-f** EDS elemental maps of Ni, Cr, and Y, respectively, acquired from the same region, showing lateral compositional uniformity of Cr and strong Y segregation to the grain boundary. **g** EDS line scan profile across the grain boundary, confirming segregation of Y to the grain boundary region.

Additional STEM-EDS measurements from two grain boundary regions in a nanocrystalline Ni-3.0Y film with a slightly smaller grain size of approximately 15 nm are provided in Figure S6 in the Supplementary Materials, with Y enrichment observed in both cases. These measurements support the idea that Y segregation to the grain boundary region is commonplace in sputtered nanocrystalline Ni-rich alloys containing this dopant .

### 3.3 Hardness data and trends

Figure 5(a) shows the hardness of the alloys as a function of Cr concentration, with three distinct datasets delineated by Y alloying levels. The results demonstrate that additions of both Cr and Y lead to substantial hardness increases, confirming that nanocrystalline solid solution strengthening and grain boundary segregation each contribute to overall alloy hardening. For the 0.5 and 1.5 at.% Y datasets, increasing Cr content initially leads to a strong increase in hardness, followed by an abrupt flattening behavior that results in a clear hardness plateau, indicating saturation of the solid solution strengthening effect. The 3 at.% Y dataset does not exhibit a well-defined plateau over the Cr concentration range investigated, yet does begin to show a decrease in the hardening slope at the higher concentrations, suggesting that a strengthening plateau is imminent. Figure 5(b) shows only the samples containing 1.5 at.% Y to clearly highlight this behavior. The hardness curve can be divided into two regimes: (1) an initial strengthening regime and (2) a subsequent saturation regime beyond a critical Cr concentration. Notably, Figure 5(a) shows that increasing Y decreases the Cr-strengthening slope in the strengthening regime, indicating a competitive effect on the marginal contribution of Cr. At the same time, increasing Y raises the initial and high-Cr hardness levels and shifts the onset of saturation to higher Cr concentrations, thereby extending the effective Cr-strengthening regime. Thus, the Ni-Cr-Y response contains both competitive and cooperative contributions. It should be noted that the use of a constant loading rate results in a slight variation in strain rate across indentation depths, so no single indentation strain rate can be quoted. Grain size is an important factor affecting the strain rate sensitivity of nanocrystalline Ni [52, 53], although composition and grain boundary chemistry may also influence the strain rate sensitivity. All samples were tested using the same loading protocol and maximum indentation depth, providing consistent experimental conditions for the hardness comparison.

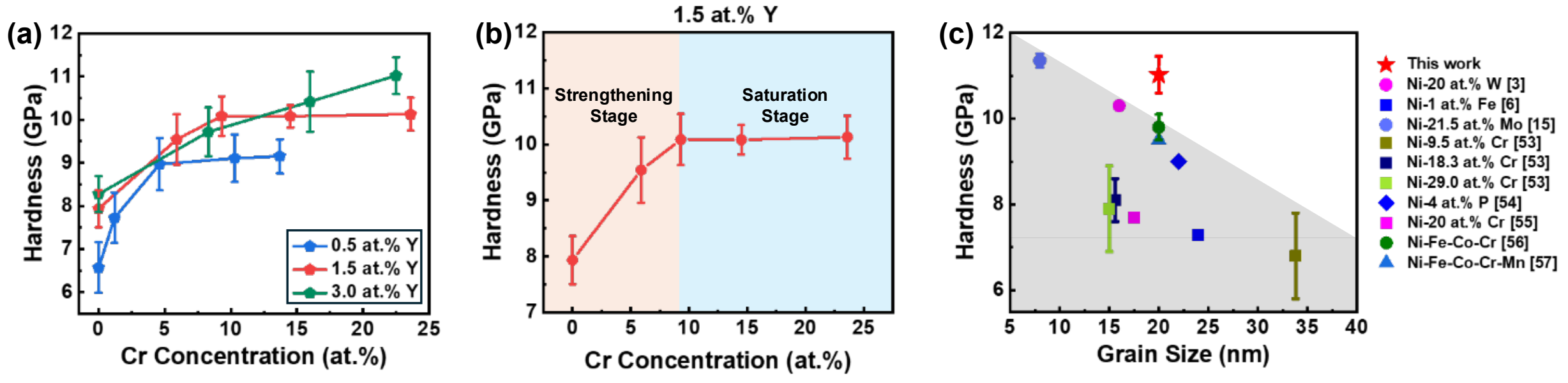

**Figure 5** Nanoindentation results demonstrate the strengthening behavior of nanocrystalline Ni-Cr-Y alloys. **a** Hardness values of all alloy films plotted as a function of Cr concentration, showing the combined effects of solid solution strengthening from Cr and grain boundary segregation from Y. **b** Representative hardness versus Cr concentration for nanocrystalline Ni-Cr-Y alloys at a fixed Y content of 1.5 at.%, showing a strengthening stage followed by a saturation stage. **c** Comparison of hardness with available literature data for single-phase nanocrystalline Ni alloys with grain sizes from 5-40 nm. **[3, 6, 15, 54–58]**.

Importantly, by combining nanocrystalline solid solution strengthening and grain boundary segregation, a maximum hardness of 11.0 GPa was achieved. Comparison with the theoretical hardness limit for Ni (~26.5 GPa, estimated based on the first-principles calculations of Liu et al. [59]) indicates that the achieved value is ~41% of this upper bound, underscoring the significant strengthening potential enabled by synergistic doping while also suggesting room for further improvement. Figure 5(c) provides a comparison with literature data for single-phase nanocrystalline Ni alloys with grain sizes near 20 nm. Studies were selected when both hardness and relevant microstructural information, particularly grain size, were available. Differences in film thickness, substrate, texture, residual stress, indentation depth, strain rate, tip calibration, and data-analysis procedures are present across these studies. The literature comparisons in this figure come from Refs. [3, 6, 15, 54–58]. The only report of comparable strength comes from Hu et al. [15], which had a much smaller grain size of 8.2 nm. Notably, the hardness measured here also exceeds that of reported nanocrystalline Ni-based high-entropy alloys, despite the latter exhibiting severe lattice distortion and additional structural features such as dense networks of nanotwins [57, 58]. As such, it can be concluded that combining solid solution strengthening and grain boundary segregation strengthening strategies can lead to outstanding strength.

The effect of Y addition can first be seen by looking at the 0 at.% Cr samples on the left side of Figure 5(a). Hardness increases from 6.6 GPa to 8.2 GPa with only the addition of 3 at.% Y. The hardness level at which the curves plateau also increases with increasing Y content, going from ~9 GPa to 11 GPa. Figure 6 summarizes these effects, by plotting the initial hardness and hardness plateau values as a function of Y concentration. For the 3 at.% Y sample set, which does not yet completely show the hardness saturation behavior (further Cr addition led to second phase precipitation), the final data point is taken as the plateau value. Increasing Y content elevates both the initial hardness and the plateau hardness to similar extents, giving increases of slightly less than 2 GPa for both. Changes to the initial hardness should come from changes to the grain boundary region, as Y is primarily found there. Dopant segregation restricts grain boundary-mediated deformation processes and therefore provides a significant strengthening contribution, consistent with previous reports [60, 61]. This strengthening has been shown to arise from a reduction in grain boundary energy that accompanies dopant segregation, such as the strong interfacial enrichment demonstrated in Figure 4, with atomistic simulations demonstrating that large solute atoms reduce grain boundary energy upon segregation to the boundary [62]. A lower-energy grain boundary occupies a deeper configurational energy minimum, such that activating the collective boundary motion required for grain sliding and rotation demands a higher applied stress, as demonstrated in Zhang et al. [25] through simulated nanoindentation on Ni-P relative to pure Ni. Vo et al. have shown computationally that this connection between grain boundary energy and flow stress is direct and quantifiable; reductions in grain boundary energy through solute doping lead to measurable increases in the strength of nanocrystalline metals [63, 64]. Molecular dynamics simulations further confirm that segregated dopants specifically improve grain boundary sliding resistance [24]. The similar increase in the plateau values suggests that this effect also comes from Y enrichment at the grain boundaries, which will be discussed in more detail in the next section.

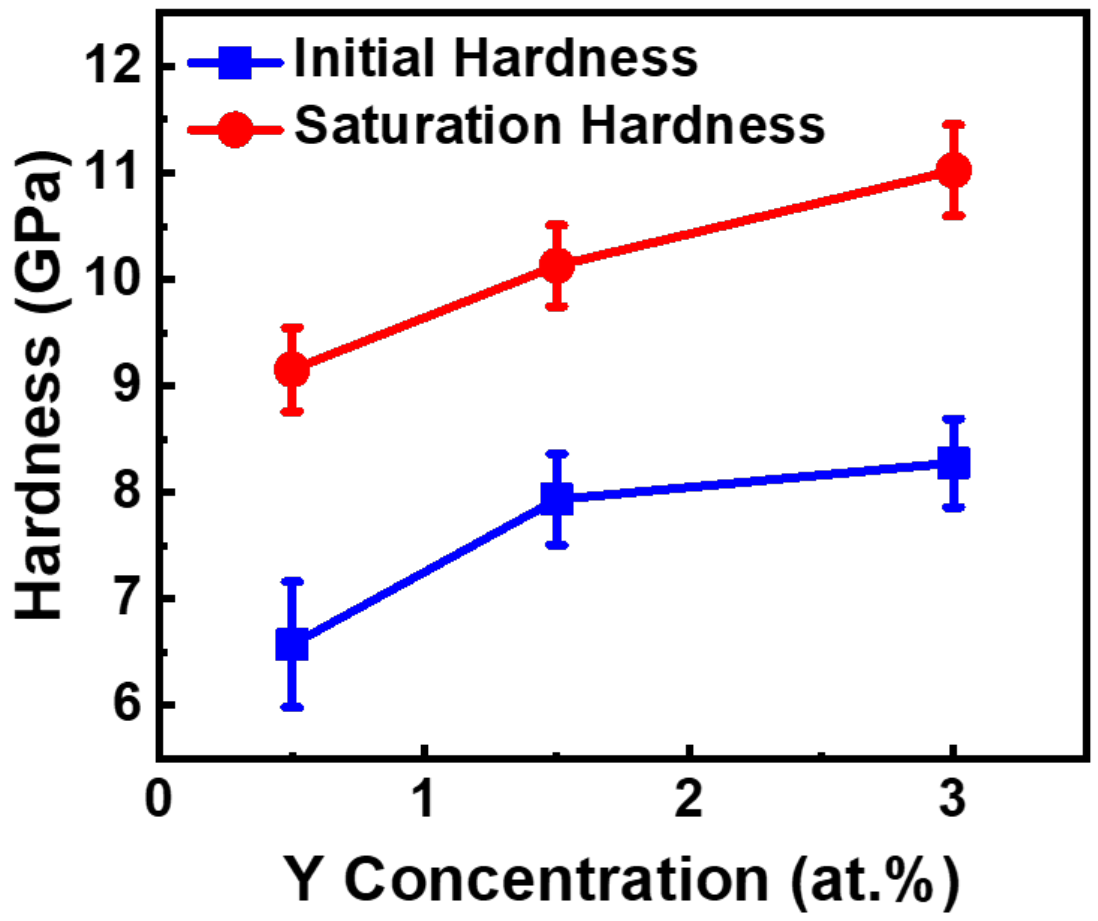

**Figure 6** The initial hardness (value when there is no Cr addition) and saturation hardness (value when Cr reaches a critical concentration after which there is no further solid solution strengthening) of nanocrystalline Ni-Cr-Y alloys as a function of Y concentration. Similar strengthening increments for both suggest a common driver being grain boundary segregation.

Y addition also affects the slope of solid solution strengthening, with more Y leading to a lower strengthening slope in Figure 5(a). Figure 7 analyzes this effect in terms of common solid solution strengthening models. In addition to the data from this study, dashed curves are provided for both the classical Fleischer model [28] and the nanocrystalline solid solution strengthening (NC-SSS) model introduced by Rupert et al. [3]. The Fleischer model describes the increase in shear strength of coarse-grained cubic alloys caused by substitutional solid solution strengthening as:

$$\Delta\tau_{Fleischer} = A \cdot G_{solvent} \cdot \left| \frac{\frac{1}{G_{solvent}} \cdot \frac{\partial G}{\partial c}}{1+\frac{1}{2}\left|\frac{1}{G_{solvent}} \cdot \frac{\partial G}{\partial c}\right|} - 3 \cdot \frac{1}{b_{solvent}} \frac{\partial b}{\partial c} \right|^{3/2} \cdot c^{\frac{1}{2}} \quad (1)$$

where $A$ is a fitting constant and can be empirically set to 0.0072 based on the literature data for coarse-grained Ni-Cr solid solution alloys [36], $G$ is the shear modulus, $b$ is the Burgers vector, and $c$ is the atomic concentration of Cr. The shear strength increase predicted by the NC-SSS model is an additional contribution to the Fleischer model as follows:

$$\Delta\tau = \Delta\tau_{Fleischer} + \Delta\tau_{nc,SS} = \Delta\tau_{Fleischer} + \frac{G_{solvent} b_{solvent}}{d} \cdot \left(\frac{1}{G_{solvent}} \frac{\partial G}{\partial c} + \frac{1}{b_{solvent}} \frac{\partial b}{\partial c}\right) \cdot c \quad (2)$$

where $d$ is the grain size. In our work, the average grain size measured by BF-TEM ($d$ = 22 nm) was used, along with a Tabor factor [65] of 3.8 for nanocrystalline Ni [66], to obtain hardness predictions as plotted in Figure 7 as the dashed lines. Notably, the experimentally observed hardness values significantly exceed the predictions of both models.

The use of grain size in the NC-SSS model assumes that this is the key dislocation obstacle spacing. Conceptually, this would mean the critical event is when a dislocation is pinned at both sides by opposite grain boundaries and must bow out to move past this configuration. This has proved powerful in many situations [30, 67, 68], yet it is also possible that other dislocation configurations are important. To remove this assumption, we introduce a very simple "modified NC-SSS" model as:

$$\Delta\tau = \Delta\tau_{Fleischer} + \frac{G_{solvent} b_{solvent}}{L_{eff}} \cdot \left(\frac{1}{G_{solvent}} \frac{\partial G}{\partial c} + \frac{1}{b_{solvent}} \frac{\partial b}{\partial c}\right) \cdot c \quad (3)$$

where $L_{eff}$ is an effective dislocation obstacle spacing that replaces the grain size term in the NC-SSS model. This formulation is consistent with the mechanistic framework of Asaro and Suresh [69], who derived that the critical shear stress for dislocation emission from grain boundary sources in nanocrystalline metals scales as $Gb/d$. While Asaro and Suresh treated grain size as the governing length scale, the present model introduces $L_{eff}$ capture the additional role of grain boundary segregation, which modifies the local emission geometry and produces an effective obstacle spacing. Application of Equation 3 to the hardening data in Figure 7 shows that the increase in strength with Cr addition, prior to the saturation point, can be well captured with a model of this form. The values of $L_{eff}$ are much smaller than the grain size and generally increase as Y content is increased. For each Y series, Equation 3 was fit by least squares to the pre-saturation data of each Y series, with $L_{eff}$ as the only adjustable parameter and all other quantities fixed at independently reported values, giving $L_{eff}$ = 1.2, 3.0 and 5.8 nm for the 0.5, 1.5 and 3.0 at.% Y series, respectively. Within a series, $L_{eff}$ is constant and does not vary with Cr concentration, as the data follows a linear trend. Equation 3 is therefore a phenomenological single-parameter description, with $L_{eff}$ being an effective quantity obtained by fitting the data rather than measured directly. One must consider all possible deformation mechanisms to understand these observations.

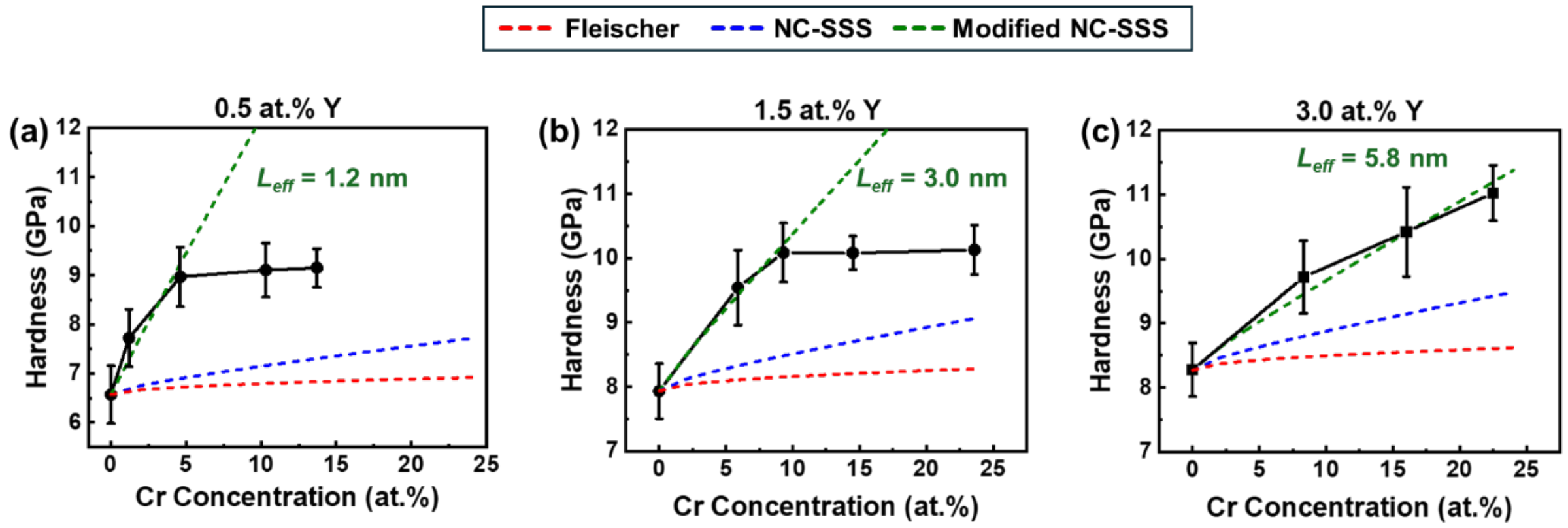

**Figure 7 a-c** Comparison of experimentally measured hardness values with predictions from the classical Fleischer model and a nanocrystalline solid solution strengthening (NC-SSS) model for samples with different Y concentrations. In all cases, the measured hardness values exceed predictions, requiring a much smaller effective obstacle spacing in a modified NC-SSS model.

As schematically illustrated in Figures 8(a)-(c), dislocation-based plasticity in nanocrystalline metals involves three primary steps: (1) dislocation emission, (2) propagation across the grain, and (3) absorption at the opposite grain boundary. Dislocation absorption is important for eventual failure (see, e.g., [70, 71]), while dislocation emission and propagation should be the key events for the initiation of plasticity. Emission is often thought to be a relatively easy event, since local stress concentrations at grain boundaries can be high and enable the dislocation to rapidly move into the grain [69]. However, the significant doping of grain boundaries here should alter this event. Reduction of grain boundary energy comes with a relaxation of local structure, which may make the emission mechanism more difficult. Since the configuration for emission is necessarily smaller than the grain size, this could explain the smaller $L_{eff}$ values shown in Figure 7.

Figures 8(d)-(f) show three proposed dislocation configurations for the emission event, one each for different levels of grain boundary doping. These configurations show schematic representations that are consistent with the fitted trend in $L_{eff}$ rather than a directly observed geometry. In all cases, the dislocations are pinned at the same location along the grain boundary, with the emission state defined from nanoscale structural features determined by the exact grain boundary types. Importantly, for the same grain with the same confining grain boundaries, one would expect the pinning locations to be consistent. Progressive doping of the boundary and associated reduction in boundary energy would mean that the restricting force on the dislocation increases. For the same pinning locations, the dislocation would become less bowed out and instead take on flatter geometries. Such behavior would increase the dislocation's effective radius of curvature radius, which corresponds directly to an increasing $L_{eff}$ values as Y concentration increases. To understand why this would affect the slope of the solid solution strengthening, one must think about the line length of the dislocations shown in Figure 8(d)-(f). The curved configuration in Figure 8(d) is longer and will therefore interact with more solute atoms. As the configuration flattens, few atoms will come in contact and the solid solution strengthening slope should be reduced, matching the trends in our hardness data in Figure 7. The configurations shown in Figures 8(d)-(f) represent a proposed interpretation of the hardness trends and fitted $L_{eff}$ values; the dislocation emission geometries were not directly observed.

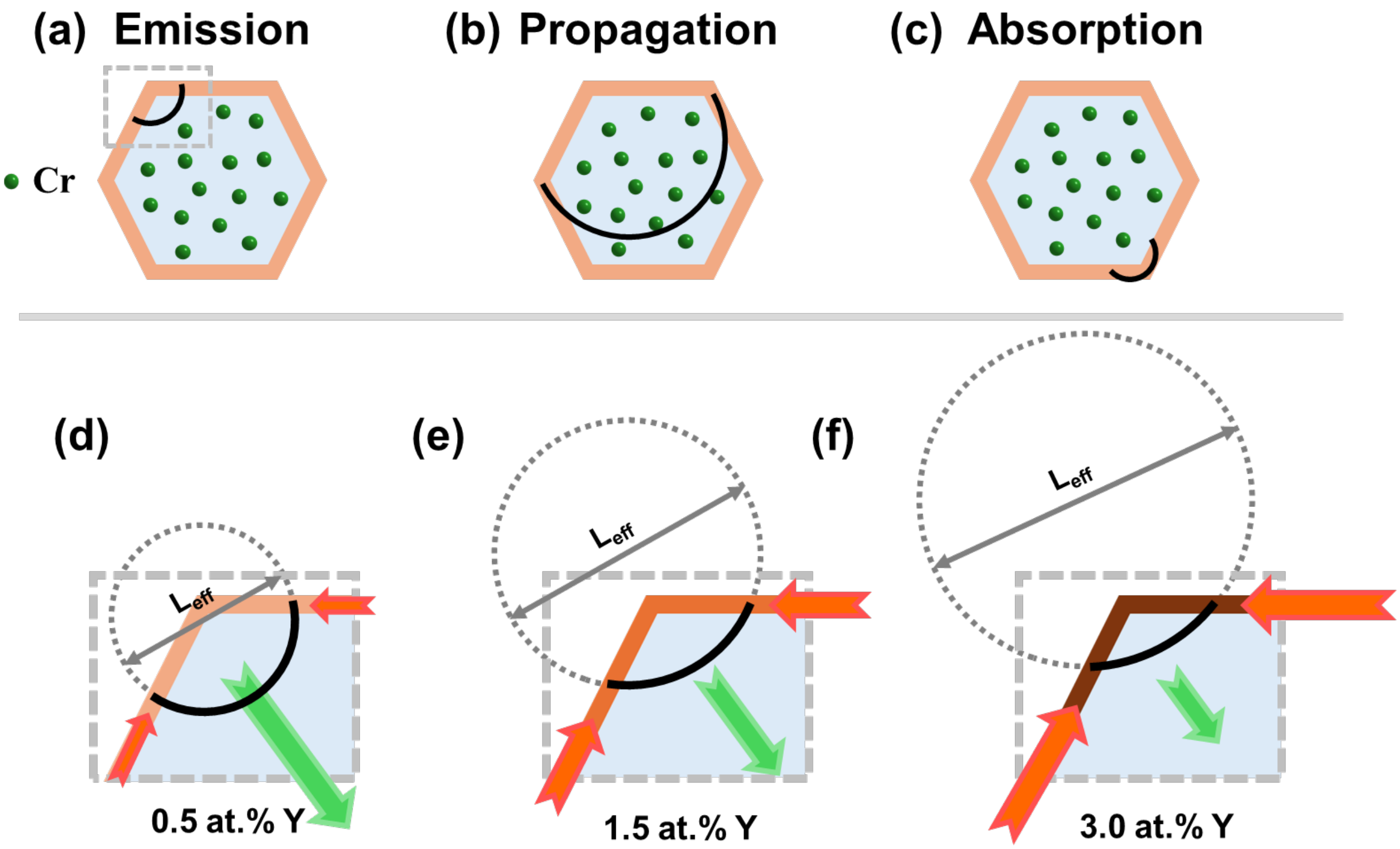

**Figure 8** Schematic illustrations of key dislocation processes in nanocrystalline alloys: **a** dislocation emission from a grain boundary nucleation site, **b** dislocation propagation across the grain interior, and **c** dislocation absorption at the opposite grain boundary. **d-f** Magnified schematics illustrating the proposed effect of Y segregation on the dislocation emission process. With increasing Y content, segregated atoms are proposed to increase the restricting force along the grain boundary, making it more difficult for the dislocation to bow into the grain interior and thereby increasing the fitted effective obstacle spacing, $L_{eff}$. A larger $L_{eff}$ corresponds to a larger curvature radius and a shorter mobile dislocation segment, which should translate to fewer interactions with solute atoms. In **d-f**, red arrows indicate the restricting force exerted by the Y-segregated grain boundary region, green arrows indicate dislocation propagation into the grain interior, and the grey arrow marks the effective dislocation obstacle spacing, $L_{eff}$.

### 3.4 Hardness saturation and shifting deformation mechanisms

Finally, the hardness saturation shown in Figure 5(b) suggests an even more dramatic change in the dominant deformation mechanism. In this regime, the strength of the alloy is no longer a function of the Cr concentration added in solid solution to the grain interior. At the same time, it is sensitive to the Y concentration at the grain boundaries. These two factors together suggest an increased importance of the grain boundary region, along with decreased importance of the lattice interior where dislocations are active. Hence, we hypothesize that the dominant deformation mechanism becomes one that primarily depends on the grain boundary state. Grain boundary-mediated mechanisms such as sliding, rotation, and grain growth become increasingly active as one approaches the Hall-Petch breakdown regime [72, 73]. While our grain sizes are slightly above the typical critical grain size for pure Ni, this exact transition may change with heavy alloying and subsequent changes to the mechanical properties.

Generally, grain boundary mechanisms are thought to be more collective in nature than dislocation-based ones, with grains shifting and moving together cooperatively. Additional cube corner nanoindentation experiments were performed on a sample within the solid solution strengthening stage (Ni-5.9Cr-1.3Y, low-Cr sample) and another within the hardness saturation stage (Ni-14.5Cr-1.5Y, high-Cr sample). This indentation geometry imposes a higher stress concentration and promotes localized plasticity [74–76]. As shown in Figures 9(a) and (e), the residual indentation impressions in both samples have pronounced pile-up regions due to the sharp indenter shape. Notably, Figures 9(b)-(d) and (f)-(h) reveal that the representative high-Cr and low-Cr samples show different degrees of plastic localization. The low-Cr sample displays a smooth and uniform pile-up surface, as evidenced by the single, well-defined peak in the line scan of Figure 9(d). The high-Cr sample exhibits markedly more heterogeneous plastic deformation; multiple surface slip steps are visible in the three-dimensional AFM image of Figure 9(f) and are marked by red arrowheads. The corresponding AFM line scan in Figure 9(h) reveals an irregular profile, in stark contrast to the smooth single peak seen in the low-Cr counterpart. These slip steps are direct evidence of plastic localization, which has been directly connected with grain boundary-mediated mechanisms in the literature. For example, Trelewicz and Schuh [76] showed that grain size reduction in nanocrystalline Ni-W eventually led to inhomogeneous shear banding near indentations that resembled the deformation of metallic glass, with visible shear steps emerging in the indentation pile-up at the finest grain sizes. The underlying physical picture is that grain boundary-mediated processes produce discrete, spatially localized displacement increments that manifest as surface undulations rather than the smooth pile-up associated with dislocation-based plasticity [74, 76, 77]. Using indentation combined with AFM, Chinh et al. [78] directly demonstrated this behavior in ultrafine-grained aluminum, where grain boundary sliding produced abrupt surface displacements at intervals matching the grain size, a pattern similar to the pile-up roughening observed here in the high-Cr sample. The transition from smooth to rough pile-up morphology observed here therefore provides morphological evidence for a shift in the operative deformation mechanism, one that is consistent with the hardness saturation behavior where the grain interior concentration is no longer of primary importance.

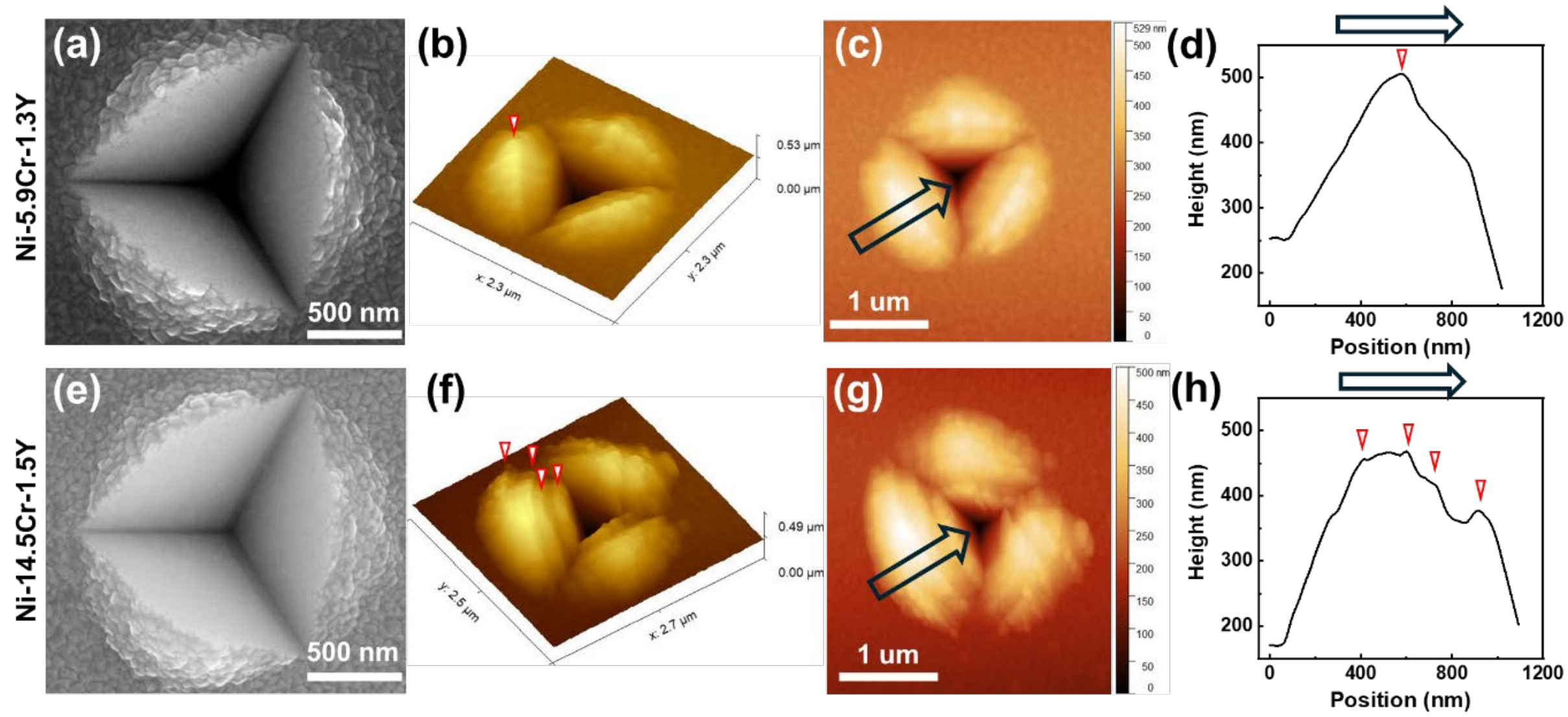

**Figure 9** Cube corner nanoindentation surface morphologies of **a-d** Ni-5.9Cr-1.3Y, which is in the regime where solid solution strengthening is occurring, and **e-h** Ni-14.5Cr-1.5Y, which is in the hardness saturation regime. **a** and **e** SEM plane-view images of the residual indents. **b** and **f** Three-dimensional AFM topographic images of the pile-up regions; the red arrowhead in **b** marks the single smooth height maximum of the low-Cr pile-up, while the red arrowheads in **f** mark multiple slip steps in the high-Cr sample, indicative of heterogeneous plasticity. **c** and **g** Two-dimensional AFM height maps of the same indent locations. **d** and **h** Height profiles extracted along the scan directions indicated in **c** and **g**, showing a single well-defined peak for the low-Cr sample and an irregular, multi-peak profile for the high-Cr sample. The higher Cr content sample exhibits a larger and more heterogeneous pile-up, indicating enhanced collective deformation that is common with grain boundary-mediated plasticity.

To provide further evidence of changing deformation mechanisms, TEM lamellae of the representative high-Cr and low-Cr samples were lifted out from the pile-up region. As shown in Figures 10(a)-(d), BF-TEM imaging confirms that both samples retain nanoscale grain structures after deformation, with no evidence of coarsening in the pile-up region. This indicates that the observed plateau behavior is not caused by deformation-induced grain growth. In the low-Cr sample, although some local grain tilting is evident, the tilted grains are very limited in number and their orientation changes remain relatively localized, without a strong common directional tendency across the pile-up region. This indicates that cooperative deformation involving multiple neighboring grains does not dominate plasticity. In contrast, the high-Cr sample exhibits a larger number of tilted grains with a more evident directional bias (examples denoted by orange arrows in Figure 10(d)), with the original columnar structure largely retained yet tilted (dashed orange lines in Figure 10(c)), suggesting that more grains participate in cooperative reorientation during pile-up formation. Such collective deformation requires coordinated strain accommodation among neighboring grains, meaning that these observations are consistent with enhanced grain-boundary-mediated plasticity in the high-Cr condition, with grain boundary sliding and grain rotation likely contributing as major accommodation mechanisms. Combined with the slip localization observed in the pile-up morphology from the cube corner indentation experiments, these TEM results support a transition in the dominant deformation mode at high Cr contents. Once the grain interior becomes sufficiently strengthened by solute addition, further intragranular dislocation activity becomes increasingly constrained and plastic deformation is accommodated more readily by grain-boundary-mediated collective processes. Such a transition provides a microstructural explanation for the observed hardness plateau. The systematic increase in saturation hardness with overall Y content can be interpreted within the same framework. If increasing Y content produces greater grain-boundary enrichment, the segregated atoms may increase resistance to local grain-boundary plasticity, requiring a higher applied stress before these mechanisms become dominant. This interpretation is consistent with the molecular dynamics simulations of Zhang et al. [25] , which showed substantially reduced grain microrotation in P-doped Ni boundaries relative to pure Ni. By the same reasoning, increasing Y content may delay the onset of hardness saturation to higher Cr concentrations.

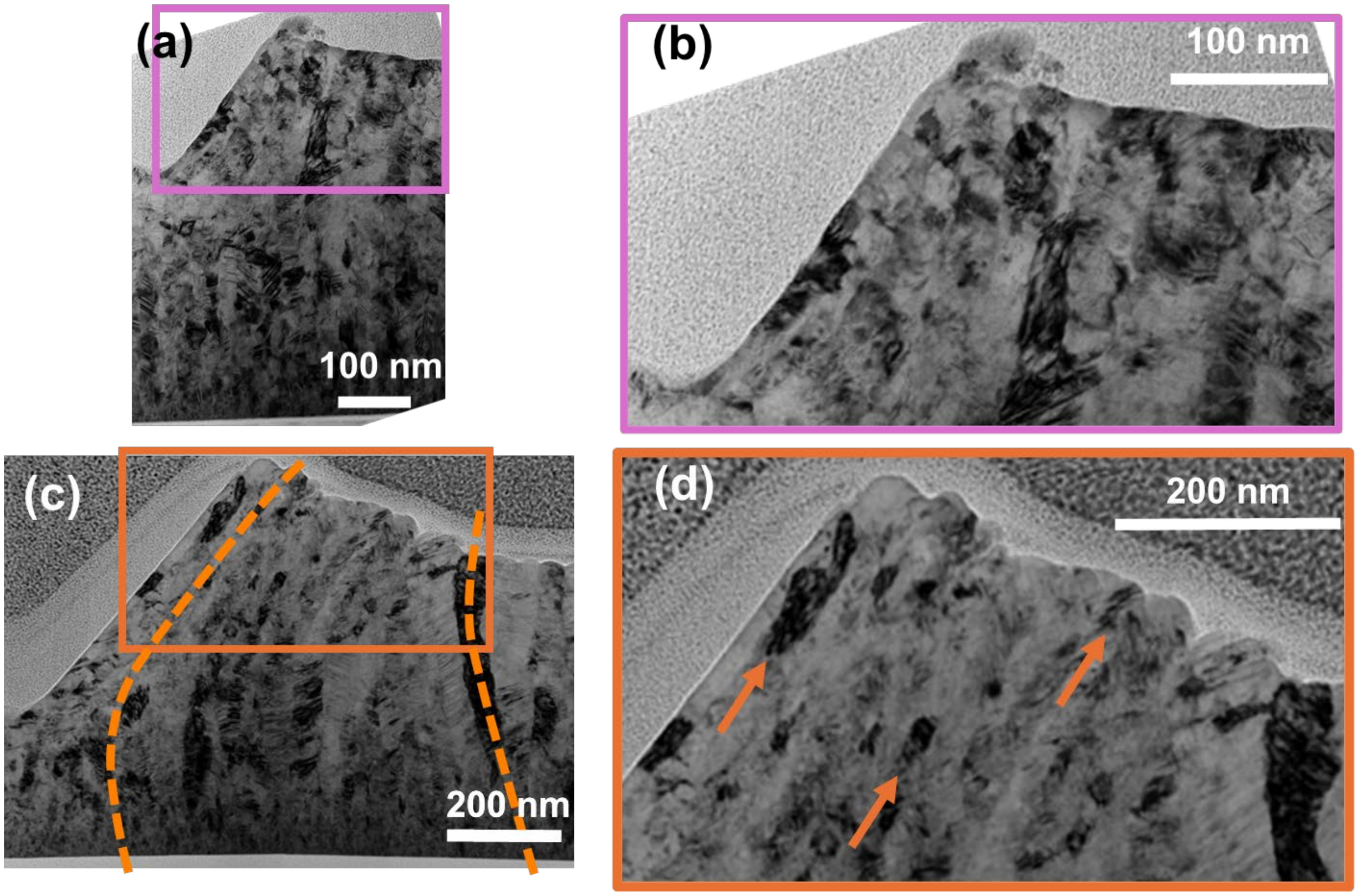

**Figure 10** BF-TEM images of the deformed pile-up region at different magnifications. **a** and **b** Low- and high-magnification views of Ni-5.9Cr-1.3Y in the pre-saturation regime; **c** and **d** low- and high-magnification views of Ni-14.5Cr-1.5Y in the hardness saturation regime. **b** and **d** correspond to the magnified views of the boxed regions indicated in **a** and **c**, respectively. Orange dashed lines in **c** show the columnar structures from the original, as-deposited microstructure while orange arrows in **d** indicate grains with pronounced tilting, consistent with cooperative grain rotation. Such features are not observed in the low-Cr sample. No grain growth was detected in either sample.

Nanotwinning may provide an additional strengthening contribution by impeding dislocation motion and modifying dislocation nucleation, propagation, and grain-boundary-mediated deformation. However, previous work on binary Ni-Cr alloys has shown that the stacking-fault energy of Ni decreases with increasing Cr concentration [79], while reduced fault energies have also been associated with increased twin frequency and reduced twin spacing in Ni-Cr films [80]. Thus, the propensity for twinning would not be expected to disappear abruptly at the onset of hardness saturation and cannot be responsible for this behavior. Nanotwinning should therefore be regarded as an additional contributor to hardening, but the present data does not suggest that it dominates compared to solid-solution strengthening and grain-boundary-mediated deformation.

The concurrent shift of the hardness saturation onset to higher Cr concentrations and the elevation of the plateau hardness with increasing Y content are consistent with a role of grain-boundary segregation in modifying resistance to collective deformation. As a proposed interpretation, Y segregation may reduce grain-boundary energy and raise the stress barrier for the activation of grain-boundary-mediated processes. In the saturation regime, segregated Y atoms may restrict the nucleation or motion of disconnections associated with grain rotation [81, 82]. Segregated Y atoms are expected to pin the disconnections in direct analogy to the pinning of lattice dislocations by bulk solutes in the grain interior. As grain boundary sliding and grain rotation become more difficult, plastic deformation from intragranular dislocation activity can persist over a larger range of Cr concentrations since the easiest deformation mechanisms "wins" and dominates plasticity. Such an interpretation is consistent with prior work showing that segregation-modified boundary structures can increase the flow stress associated with nanocrystalline plasticity and suppress grain-boundary-dominated deformation, while grain boundary processes can produce discrete surface displacements similar to those observed in the pile-up morphology discussed above [78].

# 4 Summary and Conclusions

In this study, nanocrystalline Ni-Cr-Y thin films with varying Cr and Y contents were synthesized to analyze the effects of solid solution strengthening and grain boundary segregation.

Mechanical behavior was evaluated via nanoindentation, while the deformation mechanisms were investigated using various TEM-based modalities. The following major conclusions can be drawn:

1. Cr additions enhance hardness through classical and nanocrystalline solid solution strengthening contributions, yet this effect requires a smaller effective obstacle spacing and saturates at high concentrations due to the activation of grain boundary-mediated plasticity.
2. Y additions do not significantly contribute to solid solution strengthening, while preferential Y enrichment is observed at the examined grain boundaries. The associated increase in hardness is consistent with grain boundary segregation strengthening, while the proposed effect on dislocation emission was not directly observed.
3. Increasing Y content reduces the Cr-hardening slope. Within the proposed model, this trend is interpreted as arising from segregation-induced restriction of dislocation bowing, which shortens the mobile dislocation segment and reduces the extent of dislocation-solute interactions.
4. Cube corner nanoindentation and post-indentation TEM reveal more heterogeneous pile-up formation and more extensive grain reorientation in the high-Cr saturation regime. These features are consistent with increased grain-boundary-mediated plasticity, with grain-boundary sliding and rotation as proposed mechanisms. Y segregation may increase resistance to these processes and thereby extend the effective Cr-strengthening range.
5. The highest strength ternary alloy composition achieves a maximum hardness of 11.0 GPa, among the highest values reported for Ni-based single phase nanocrystalline alloys to date and the highest for this grain size. This result underscores the potential of combining nanocrystalline grain structures with targeted grain boundary segregation to realize extreme mechanical performance in conventional alloy systems.

As a whole, the results shown here demonstrate that solid solution strengthening and grain boundary segregation do not operate independently in multicomponent nanocrystalline alloys. Instead, their interaction first governs the activation of dislocation-mediated plasticity and then the eventual transition to grain boundary-dominated mechanisms. Segregation alters grain boundary-mediated dislocation processes, which in turn influence the effectiveness of lattice strengthening by modifying the geometry of dislocation emission. Moreover, segregation suppresses grain boundary sliding and other boundary-controlled deformation modes, allowing solid solution strengthening to remain active over a broader compositional range. Together, these competitive and cooperative contributions produce hardness levels and strength-solute scaling trends in the Ni-Cr-Y system that are larger than existing theoretical predictions, demonstrating a synergistic overall effect. These findings highlight the critical importance of coordinating solute distribution and interface chemistry in the design of high-performance nanocrystalline alloys.

## 5 Statements and Declarations

**Competing Interests:** The authors declare that they have no known competing financial interests or personal relationships that could have appeared to influence the work reported in this paper.

## 6 Data availability

The datasets generated during and/or analyzed during the current study are available from the corresponding author on reasonable request.

## 7 Acknowledgements

This work was supported by the National Science Foundation under the Grant Number DMR-2502676. Trelewicz acknowledges the same collaborative award under grant number DMR-2310306. The authors acknowledge the use of facilities and instrumentation at the UC Irvine Materials Research Institute (IMRI), which is supported in part by the National Science Foundation through the UC Irvine Materials Research Science and Engineering Center (DMR-2011967). SEM and FIB work were performed using instrumentation funded in part by the National Science Foundation Center for Chemistry at the Space-Time Limit (CHE-0802913).

## References

1. Koch CC (2007) Structural nanocrystalline materials: an overview. J Mater Sci 42:1403–1414. https://doi.org/10.1007/s10853-006-0609-3

2. Chookajorn T, Murdoch HA, Schuh CA (2012) Design of Stable Nanocrystalline Alloys. Science 337:951–954. https://doi.org/10.1126/science.1224737

3. Rupert TJ, Trenkle JC, Schuh CA (2011) Enhanced solid solution effects on the strength of nanocrystalline alloys. Acta Materialia 59:1619–1631. https://doi.org/10.1016/j.actamat.2010.11.026

4. Dao M, Lu L, Asaro R, et al (2007) Toward a quantitative understanding of mechanical behavior of nanocrystalline metals. Acta Materialia 55:4041–4065. https://doi.org/10.1016/j.actamat.2007.01.038

5. Henager CH (2024) Reversing inverse Hall-Petch and direct computation of Hall-Petch coefficients. Acta Materialia 265:119627. https://doi.org/10.1016/j.actamat.2023.119627

6. Shen T, Schwarz R, Feng S, et al (2007) Effect of solute segregation on the strength of nanocrystalline alloys: Inverse Hall–Petch relation. Acta Materialia 55:5007–5013. https://doi.org/10.1016/j.actamat.2007.05.018

7. Wang L, Zhang Y, Zeng Z, et al (2022) Tracking the sliding of grain boundaries at the atomic scale. Science 375:1261–1265. https://doi.org/10.1126/science.abm2612

8. Farkas D, Frøseth A, Van Swygenhoven H (2006) Grain boundary migration during room temperature deformation of nanocrystalline Ni. Scripta Materialia 55:695–698. https://doi.org/10.1016/j.scriptamat.2006.06.032

9. Wang L, Teng J, Liu P, et al (2014) Grain rotation mediated by grain boundary dislocations in nanocrystalline platinum. Nat Commun 5:4402. https://doi.org/10.1038/ncomms5402

10. Frøseth AG, Derlet PM, Van Swygenhoven H (2004) Dislocations emitted from nanocrystalline grain boundaries: nucleation and splitting distance. Acta Materialia 52:5863–5870. https://doi.org/10.1016/j.actamat.2004.09.001

11. Van Swygenhoven H, Derlet PM, Hasnaoui A (2002) Atomic mechanism for dislocation emission from nanosized grain boundaries. Phys Rev B 66:024101. https://doi.org/10.1103/PhysRevB.66.024101

12. Turlo V, Rupert TJ (2018) Grain boundary complexions and the strength of nanocrystalline metals: Dislocation emission and propagation. Acta Materialia 151:100–111. https://doi.org/10.1016/j.actamat.2018.03.055

13. Van Swygenhoven H, Derlet PM, Frøseth AG (2006) Nucleation and propagation of dislocations in nanocrystalline fcc metals. Acta Materialia 54:1975–1983. https://doi.org/10.1016/j.actamat.2005.12.026

14. Kondo S, Mitsuma T, Shibata N, Ikuhara Y (2016) Direct observation of individual dislocation interaction processes with grain boundaries. Sci Adv 2:e1501926. https://doi.org/10.1126/sciadv.1501926

15. Hu J, Shi YN, Sauvage X, et al (2017) Grain boundary stability governs hardening and softening in extremely fine nanograined metals. Science 355:1292–1296. https://doi.org/10.1126/science.aal5166

16. Schiøtz J, Di Tolla FD, Jacobsen KW (1998) Softening of nanocrystalline metals at very small grain sizes. Nature 391:561–563. https://doi.org/10.1038/35328

17. Luo J (2024) Grain boundary segregation models for high-entropy alloys: Theoretical formulation and application to elucidate high-entropy grain boundaries. Journal of Applied Physics 135:165303. https://doi.org/10.1063/5.0200669

18. Borovikov V, Mendelev MI, King AH (2018) Solute effects on interfacial dislocation emission in nanomaterials: Nucleation site competition and neutralization. Scripta Materialia 154:12–15. https://doi.org/10.1016/j.scriptamat.2018.05.011

19. Borovikov V, Mendelev MI, King AH (2017) Effects of solutes on dislocation nucleation from grain boundaries. International Journal of Plasticity 90:146–155. https://doi.org/10.1016/j.ijplas.2016.12.009

20. Qian L, Zhang J, Yang W, et al (2025) Maintaining Grain Boundary Segregation-Induced Strengthening Effect in Extremely Fine Nanograined Metals. Nano Lett 25:5493–5501. https://doi.org/10.1021/acs.nanolett.5c01032

21. Zhuang Q, Liang D, Luo J, et al (2025) Dual-Nano Composite Design with Grain Boundary Segregation for Enhanced Strength and Plasticity in CoCrNi-CuZr Thin Films. Nano Lett 25:691–698. https://doi.org/10.1021/acs.nanolett.4c04755

22. Picard E-A, Sansoz F (2022) Ni solute segregation and associated plastic deformation mechanisms into random FCC Ag, BCC Nb and HCP Zr polycrystals. Acta Materialia 240:118367. https://doi.org/10.1016/j.actamat.2022.118367

23. Zuo J, Nakata T, Xu C, et al (2022) Effect of grain boundary segregation on microstructure and mechanical properties of ultra-fine grained Mg–Al–Ca–Mn alloy wires. Materials Science and Engineering: A 848:143423. https://doi.org/10.1016/j.msea.2022.143423

24. Millett PC, Selvam RP, Saxena A (2006) Improving grain boundary sliding resistance with segregated dopants. Materials Science and Engineering: A 431:92–99. https://doi.org/10.1016/j.msea.2006.05.074

25. Zhang Y, Tucker GJ, Trelewicz JR (2017) Stress-assisted grain growth in nanocrystalline metals: Grain boundary mediated mechanisms and stabilization through alloying. Acta Materialia 131:39–47. https://doi.org/10.1016/j.actamat.2017.03.060

26. Masuda T, Sauvage X, Hirosawa S, Horita Z (2020) Achieving highly strengthened Al–Cu–Mg alloy by grain refinement and grain boundary segregation. Materials Science and Engineering: A 793:139668. https://doi.org/10.1016/j.msea.2020.139668

27. Guo T, Huang P, Xu KW, et al (2016) Solid solution effects on hardness and strain rate sensitivity of nanocrystalline NiFe alloy. Materials Science and Engineering: A 676:501–505. https://doi.org/10.1016/j.msea.2016.08.120

28. Schuh CA, Nieh TG, Iwasaki H (2003) The effect of solid solution W additions on the mechanical properties of nanocrystalline Ni. Acta Materialia 51:431–443. https://doi.org/10.1016/S1359-6454(02)00427-5

29. Fleischer RL (1963) Substitutional solution hardening. Acta Metallurgica 11:203–209. https://doi.org/10.1016/0001-6160(63)90213-X

30. Kim K, Park S, Kim T, et al (2022) Mechanical, electrical properties and microstructures of combinatorial Ni-Mo-W alloy films. Journal of Alloys and Compounds 919:165808. https://doi.org/10.1016/j.jallcom.2022.165808

31. Zhang XF, Fujita T, Pan D, et al (2010) Influences of grain size and grain boundary segregation on mechanical behavior of nanocrystalline Ni. Materials Science and Engineering: A 527:2297–2304. https://doi.org/10.1016/j.msea.2009.12.005

32. Atwater MA, Darling KA (2012) A Visual Library of Stability in Binary Metallic Systems: The Stabilization of Nanocrystalline Grain Size by Solute Addition: Part 1: Defense Technical Information Center, Fort Belvoir, VA

33. Okamoto H, Schlesinger ME, Mueller EM (2016) Cr (Chromium) Binary Alloy Phase Diagrams. In: Alloy Phase Diagrams. ASM International, pp 281–297

34. Baker H (1998) Properties of Metals. In: Davis JR (ed) Metals Handbook Desk Edition, 2nd ed. ASM International, pp 114–121

35. Mishima Y, Ochiai S, Hamao N, et al (1986) Solid Solution Hardening of Nickel —Role of Transition Metal and B-subgroup Solutes—. Transactions of the Japan Institute of Metals 27:656–664. https://doi.org/10.2320/matertrans1960.27.656

36. Pelloux RMN, Grant NJ (1960) Solid solutions and second phase strengthening of nickel alloys at high and low temperatures. Trans Metall Soc AIME 218:232–237

37. Darling KA, Kecskes LJ, Atwater M, et al (2013) Thermal stability of nanocrystalline nickel with yttrium additions. J Mater Res 28:1813–1819. https://doi.org/10.1557/jmr.2013.9

38. Palumbo G, Thorpe SJ, Aust KT (1990) On the contribution of triple junctions to the structure and properties of nanocrystalline materials. Scripta Metallurgica et Materialia 24:1347–1350. https://doi.org/10.1016/0956-716X(90)90354-J

39. Kong J, Haché MJR, Tam J, et al (2022) On the extrinsic Hall-Petch to inverse Hall-Petch transition in nanocrystalline Ni-Co electrodeposits. Scripta Materialia 218:114799. https://doi.org/10.1016/j.scriptamat.2022.114799

40. Thornton JA (1975) Influence of substrate temperature and deposition rate on structure of thick sputtered Cu coatings. Journal of Vacuum Science and Technology 12:830–835. https://doi.org/10.1116/1.568682

41. Giannuzzi LA, Drown JL, Brown SR, et al (1998) Applications of the FIB lift-out technique for TEM specimen preparation. Microsc Res Tech 41:285–290. https://doi.org/10.1002/(SICI)1097-0029(19980515)41:4%3C285::AID-JEMT1%3E3.0.CO;2-Q

42. Chen J, Bull SJ (2006) On the relationship between plastic zone radius and maximum depth during nanoindentation. Surface and Coatings Technology 201:4289–4293. https://doi.org/10.1016/j.surfcoat.2006.08.099

43. Oliver WC, Pharr GM (2004) Measurement of hardness and elastic modulus by instrumented indentation: Advances in understanding and refinements to methodology. J Mater Res 19:3–20. https://doi.org/10.1557/jmr.2004.19.1.3

44. Koenig TR, Rao Z, Chason E, et al (2021) The microstructural and stress evolution in sputter deposited Ni thin films. Surface and Coatings Technology 412:126973. https://doi.org/10.1016/j.surfcoat.2021.126973

45. Zhang Y, Zhang Z, Yao W, Liang X (2022) Microstructure, mechanical properties and corrosion resistance of high-level hard Nb-Ta-W and Nb-Ta-W-Hf multi-principal element alloy thin films. Journal of Alloys and Compounds 920:166000. https://doi.org/10.1016/j.jallcom.2022.166000

46. Balogh L, Ribárik G, Ungár T (2006) Stacking faults and twin boundaries in fcc crystals determined by x-ray diffraction profile analysis. Journal of Applied Physics 100:023512. https://doi.org/10.1063/1.2216195

47. Vegard L (1921) Die Konstitution der Mischkristalle und die Raumfüllung der Atome. Zeitschrift für Physik 5:17–26. https://doi.org/10.1007/BF01349680

48. Fernandes CM, Guisbiers G, Pereira S, et al (2009) Annealing Ni nanocrystalline on WC–Co. Journal of Alloys and Compounds 482:131–136. https://doi.org/10.1016/j.jallcom.2009.04.106

49. Li JX, Shi Y-N, You ZS, Li XY (2021) Tensile strain induced texture evolution in a Ni–Mo alloy with extremely fine nanotwinned columnar grains. Materials Science and Engineering: A 812:141108. https://doi.org/10.1016/j.msea.2021.141108

50. Thompson GB, Banerjee R, Zhang XD, et al (2002) Chemical ordering and texture in Ni–25 at% Al thin films. Acta Materialia 50:643–651. https://doi.org/10.1016/S1359-6454(01)00373-1

51. Velasco L, Hodge AM (2017) Growth twins in high stacking fault energy metals: Microstructure, texture and twinning. Materials Science and Engineering: A 687:93–98. https://doi.org/10.1016/j.msea.2017.01.065

52. Maier V, Durst K, Mueller J, et al (2011) Nanoindentation strain-rate jump tests for determining the local strain-rate sensitivity in nanocrystalline Ni and ultrafine-grained Al. J Mater Res 26:1421–1430. https://doi.org/10.1557/jmr.2011.156

53. Schwaiger R, Moser B, Dao M, et al (2003) Some critical experiments on the strain-rate sensitivity of nanocrystalline nickel. Acta Materialia 51:5159–5172. https://doi.org/10.1016/S1359-6454(03)00365-3

54. Tillmann W, Kokalj D, Stangier D, et al (2018) Influence of Cr-Content on the thermoelectric and mechanical properties of NiCr thin film thermocouples synthesized on thermally sprayed Al2O3. Thin Solid Films 663:148–158. https://doi.org/10.1016/j.tsf.2018.08.023

55. Bikmukhametov I, Gupta A, Koenig TR, et al (2023) Consequences of solute partitioning on hardness in stabilized nanocrystalline alloys. Materials Science and Engineering: A 875:145113. https://doi.org/10.1016/j.msea.2023.145113

56. Petley V, Sathishkumar S, Thulasi Raman KH, et al (2015) Microstructural and mechanical characteristics of Ni–Cr thin films. Materials Research Bulletin 66:59–64. https://doi.org/10.1016/j.materresbull.2015.02.002

57. Nagy P, Rohbeck N, Roussely G, et al (2020) Processing and characterization of a multibeam sputtered nanocrystalline CoCrFeNi high-entropy alloy film. Surface and Coatings Technology 386:125465. https://doi.org/10.1016/j.surfcoat.2020.125465

58. Wang Z, Wang C, Zhao Y-L, et al (2020) High hardness and fatigue resistance of CoCrFeMnNi high entropy alloy films with ultrahigh-density nanotwins. International Journal of Plasticity 131:102726. https://doi.org/10.1016/j.ijplas.2020.102726

59. Liu Y-L, Zhang Y, Zhou H-B, et al (2008) Theoretical strength and charge redistribution of fcc Ni in tension and shear. J Phys: Condens Matter 20:335216. https://doi.org/10.1088/0953-8984/20/33/335216

60. Wang Y, Qi Y, He T, Feng M (2023) Grain refinement induced by grain boundary segregation in FeNiCrCoCu high-entropy alloys using molecular dynamics simulation of nanoindentation. Materials Chemistry and Physics 310:128489. https://doi.org/10.1016/j.matchemphys.2023.128489

61. Babicheva RI, Dmitriev SV, Bai L, et al (2016) Effect of grain boundary segregation on the deformation mechanisms and mechanical properties of nanocrystalline binary aluminum alloys. Computational Materials Science 117:445–454. https://doi.org/10.1016/j.commatsci.2016.02.013

62. Millett PC, Selvam RP, Bansal S, Saxena A (2005) Atomistic simulation of grain boundary energetics – Effects of dopants. Acta Materialia 53:3671–3678. https://doi.org/10.1016/j.actamat.2005.04.031

63. Vo NQ, Schäfer J, Averback RS, et al (2011) Reaching theoretical strengths in nanocrystalline Cu by grain boundary doping. Scripta Materialia 65:660–663. https://doi.org/10.1016/j.scriptamat.2011.06.048

64. Vo NQ, Averback RS, Bellon P, Caro A (2008) Limits of hardness at the nanoscale: Molecular dynamics simulations. Phys Rev B 78:241402. https://doi.org/10.1103/PhysRevB.78.241402

65. Tabor D (2000) The Hardness of Metals. OUP Oxford

66. Dalla Torre F, Van Swygenhoven H, Victoria M (2002) Nanocrystalline electrodeposited Ni: microstructure and tensile properties. Acta Materialia 50:3957–3970. https://doi.org/10.1016/S1359-6454(02)00198-2

67. Asaro RJ, Krysl P, Kad B (2003) Deformation mechanism transitions in nanoscale fcc metals. Philosophical Magazine Letters 83:733–743. https://doi.org/10.1080/09500830310001614540

68. Tang W, Herbert EG, Anand A, et al (2026) Use of friction stir processing to synthesize nanocrystalline, grain boundary segregating Fe-Ti alloys. Materials Science and Engineering: A 959:150050. https://doi.org/10.1016/j.msea.2026.150050

69. Asaro RJ, Suresh S (2005) Mechanistic models for the activation volume and rate sensitivity in metals with nanocrystalline grains and nano-scale twins. Acta Materialia 53:3369–3382. https://doi.org/10.1016/j.actamat.2005.03.047

70. Bitzek E, Brandl C, Weygand D, et al (2009) Atomistic simulation of a dislocation shear loop interacting with grain boundaries in nanocrystalline aluminium. Modelling Simul Mater Sci Eng 17:055008. https://doi.org/10.1088/0965-0393/17/5/055008

71. Pan Z, Rupert TJ (2014) Damage nucleation from repeated dislocation absorption at a grain boundary. Computational Materials Science 93:206–209. https://doi.org/10.1016/j.commatsci.2014.07.008

72. Guo D, Song S, Luo R, et al (2018) Grain Boundary Sliding and Amorphization are Responsible for the Reverse Hall-Petch Relation in Superhard Nanocrystalline Boron Carbide. Phys Rev Lett 121:145504. https://doi.org/10.1103/PhysRevLett.121.145504

73. Romanov AE, Kolesnikova AL, Ovid'ko IA, Aifantis EC (2009) Disclinations in nanocrystalline materials: Manifestation of the relay mechanism of plastic deformation. Materials Science and Engineering: A 503:62–67. https://doi.org/10.1016/j.msea.2008.05.053

74. Duan FH, Naunheim Y, Schuh CA, Li Y (2021) Breakdown of the Hall-Petch relationship in extremely fine nanograined body-centered cubic Mo alloys. Acta Materialia 213:116950. https://doi.org/10.1016/j.actamat.2021.116950

75. Trelewicz JR, Schuh CA (2008) The Hall–Petch breakdown at high strain rates: Optimizing nanocrystalline grain size for impact applications. Applied Physics Letters 93:171916. https://doi.org/10.1063/1.3000655

76. Trelewicz JR, Schuh CA (2007) The Hall–Petch breakdown in nanocrystalline metals: A crossover to glass-like deformation. Acta Materialia 55:5948–5958. https://doi.org/10.1016/j.actamat.2007.07.020

77. Shan Z, Stach EA, Wiezorek JMK, et al (2004) Grain Boundary-Mediated Plasticity in Nanocrystalline Nickel. Science 305:654–657. https://doi.org/10.1126/science.1098741

78. Chinh NQ, Szommer P, Horita Z, Langdon TG (2006) Experimental Evidence for Grain-Boundary Sliding in Ultrafine-Grained Aluminum Processed by Severe Plastic Deformation. Advanced Materials 18:34–39. https://doi.org/10.1002/adma.200501232

79. Li Z, Xu B, Zhao S, et al (2026) Strength-ductility synergy in Ni-Cr binary alloys: critical role of stacking fault energy. Materials Science and Engineering: A 960:150056. https://doi.org/10.1016/j.msea.2026.150056

80. Yazdani MH, Liang A, Maldonado Otero AJ, et al (2026) Composition-driven nanotwin engineering in sputtered Ni-Fe and Ni-Cr films: linking fault energetics to twin thickness. Acta Materialia 306:121890. https://doi.org/10.1016/j.actamat.2025.121890

81. Tian Y, Gong X, Xu M, et al (2024) Grain rotation mechanisms in nanocrystalline materials: Multiscale observations in Pt thin films. Science 386:49–54. https://doi.org/10.1126/science.adk6384

82. Sun XY, Fressengeas C, Taupin V, et al (2018) Disconnections, dislocations and generalized disclinations in grain boundary ledges. International Journal of Plasticity 104:134–146. https://doi.org/10.1016/j.ijplas.2018.02.003

## Supplementary Materials

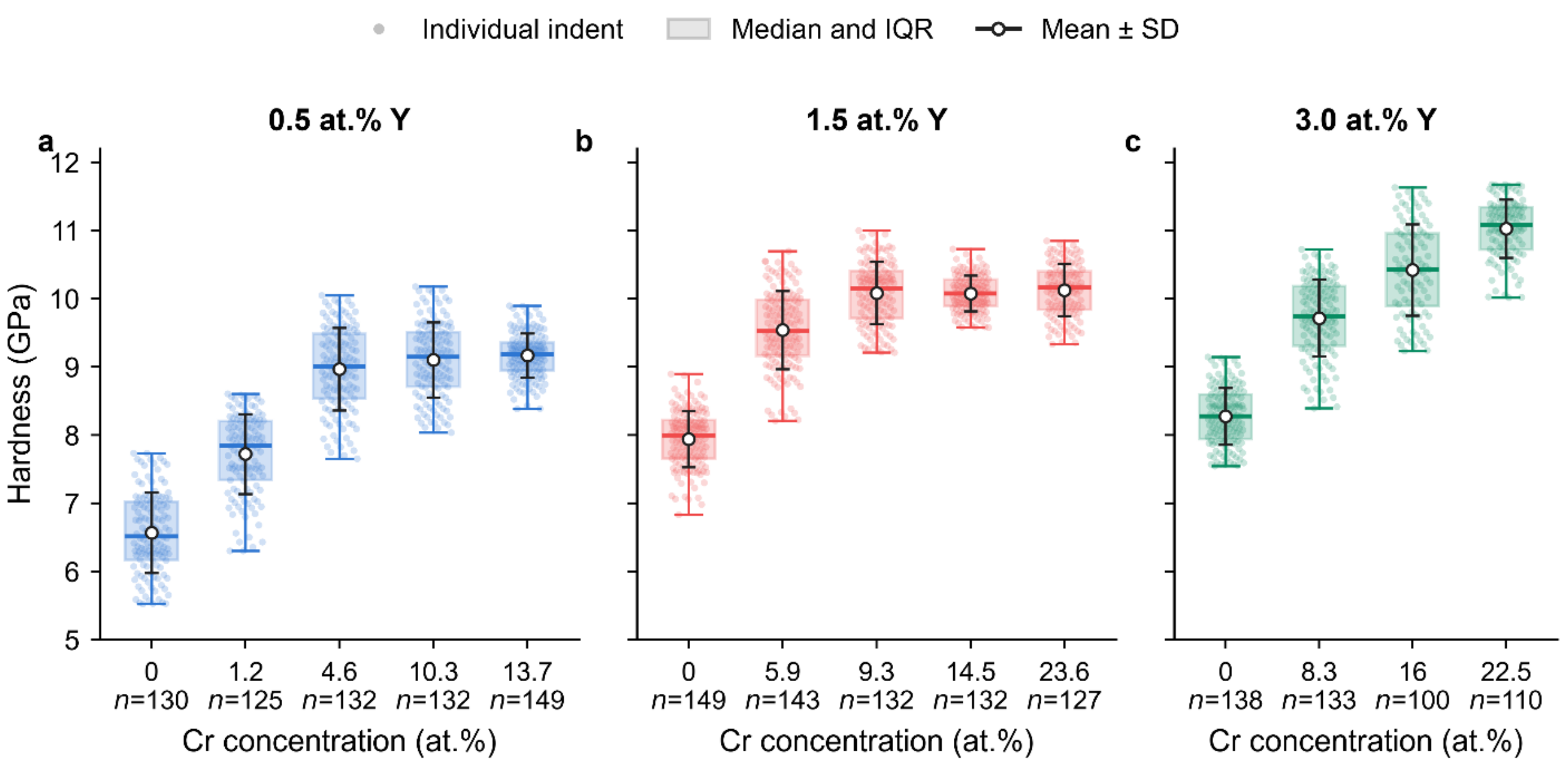

**Figure S1** Distributions of the retained individual nanoindentation hardness measurements for all investigated compositions. Individual points represent separate indentations, and the boxes indicate the median and interquartile range. The number of retained measurements ranges from 100 to 149 for each composition.

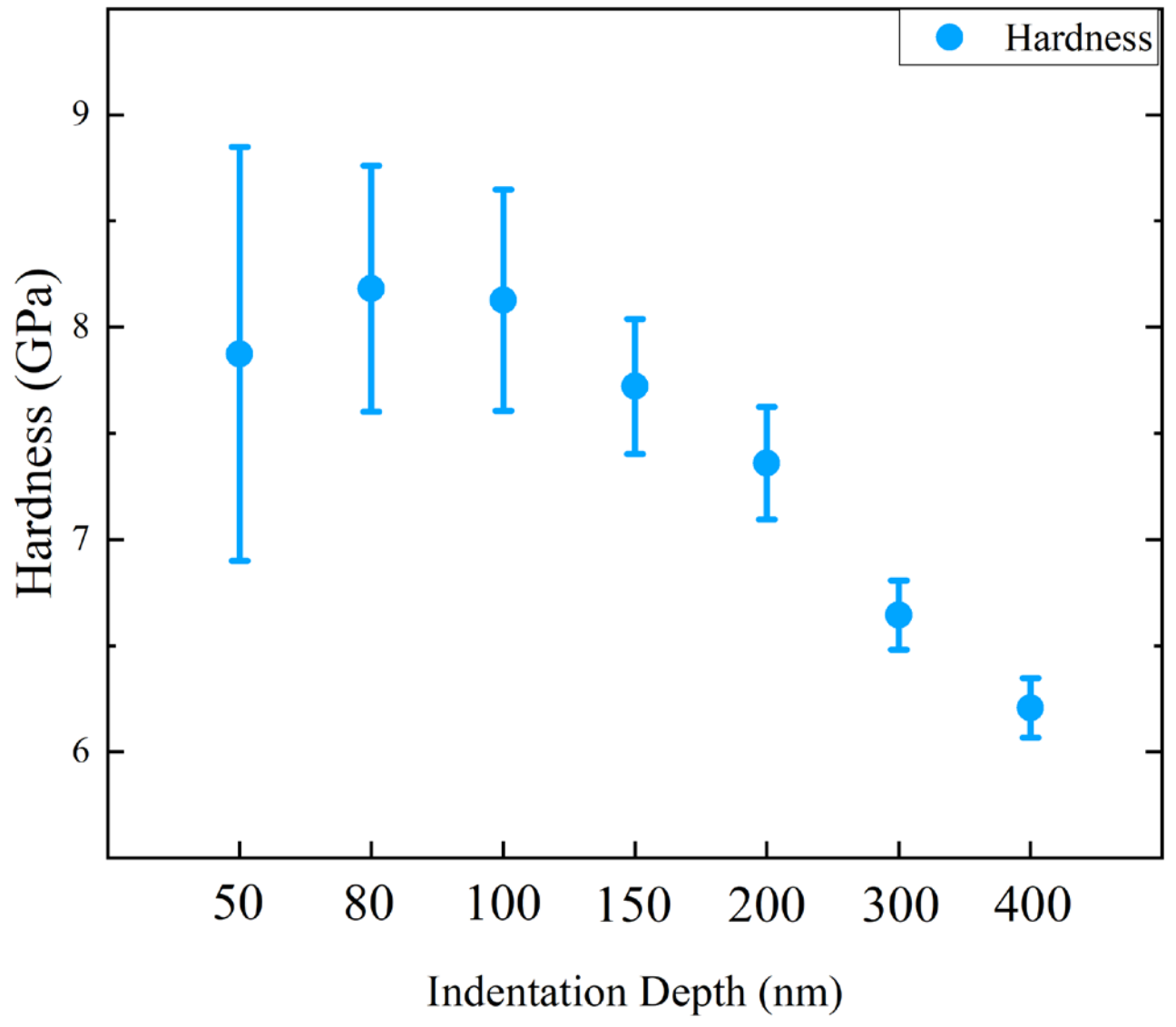

**Figure S2** Indentation-depth dependence of the hardness measured for a separately fabricated Ni-2.0Y film with a thickness and quartz substrate comparable to those of the films investigated in the main study. Data points represent the mean hardness, and error bars indicate one standard deviation. The hardness remains constant within experimental scatter between 50 and 100 nm and decreases systematically at greater indentation depths.

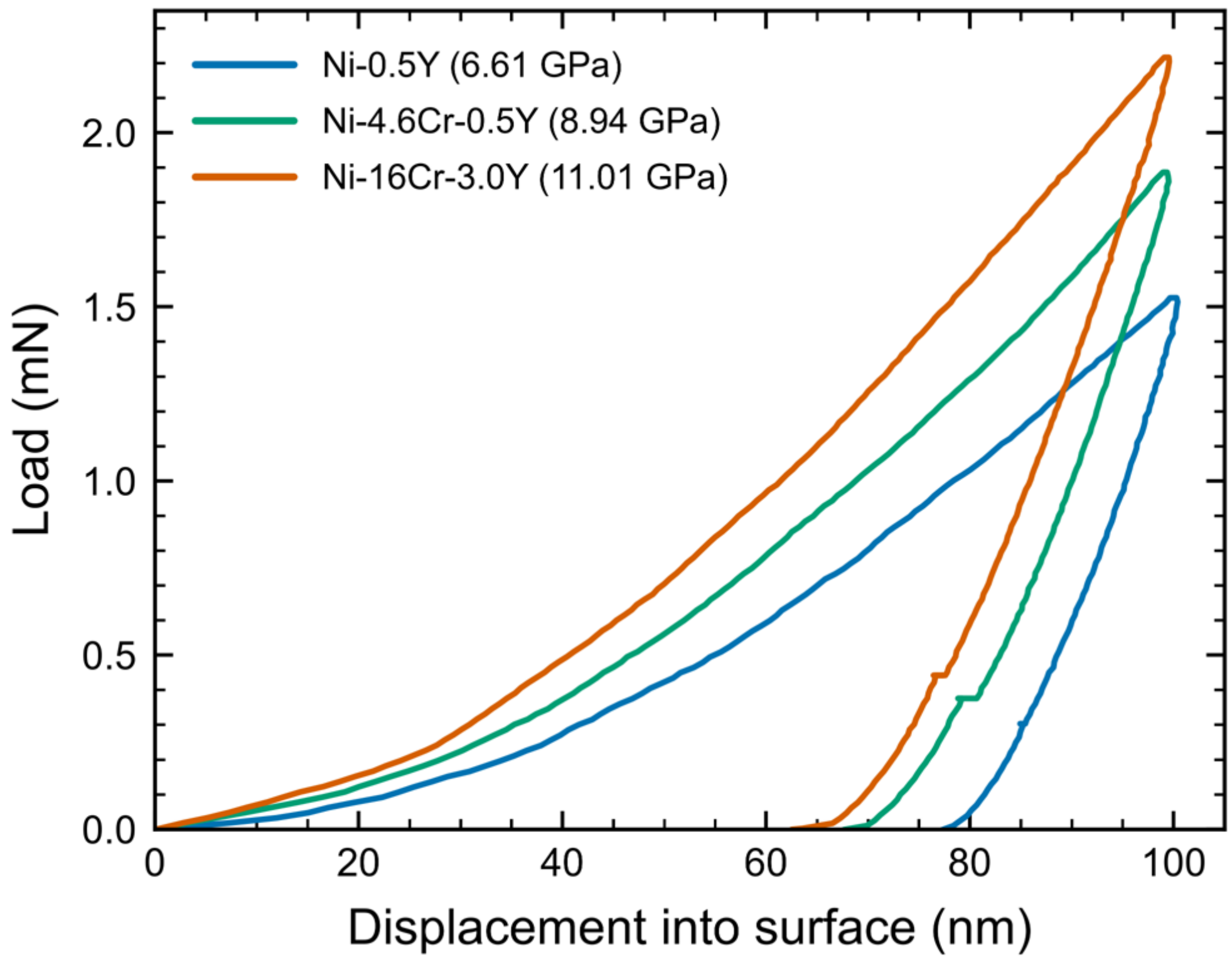

**Figure S3** Representative nanoindentation load-displacement curves for Ni-0.5Y, Ni-4.6Cr-0.5Y, and Ni-16Cr-3.0Y, selected to represent low-, intermediate-, and high-hardness alloys, respectively. The values in parentheses indicate the corresponding mean hardness.

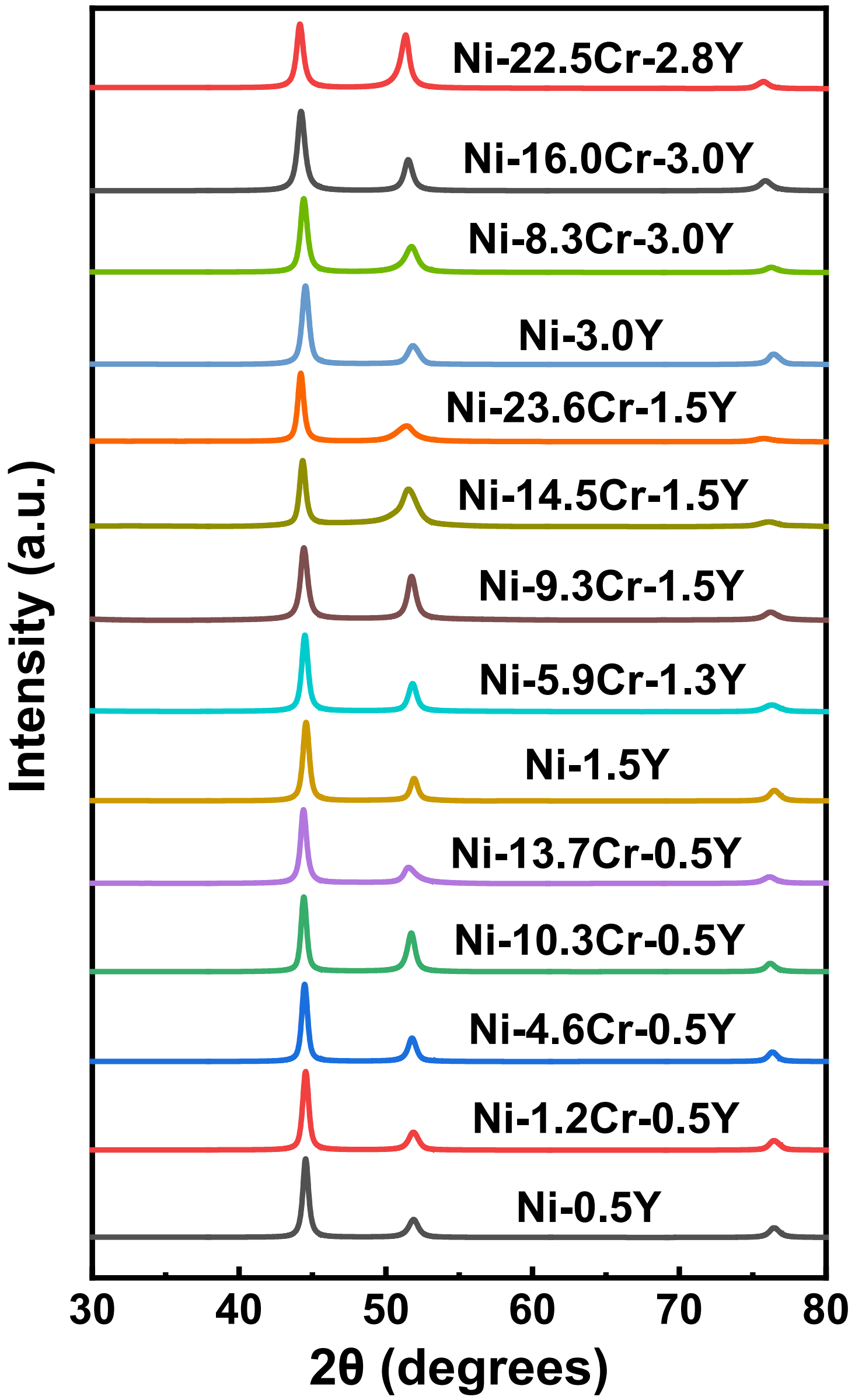

**Figure S4** Complete grazing-incidence X-ray diffraction patterns for all investigated Ni-Cr-Y compositions. The patterns are vertically offset for clarity and show the single-phase FCC Ni-rich structure without detectable secondary-phase reflections.

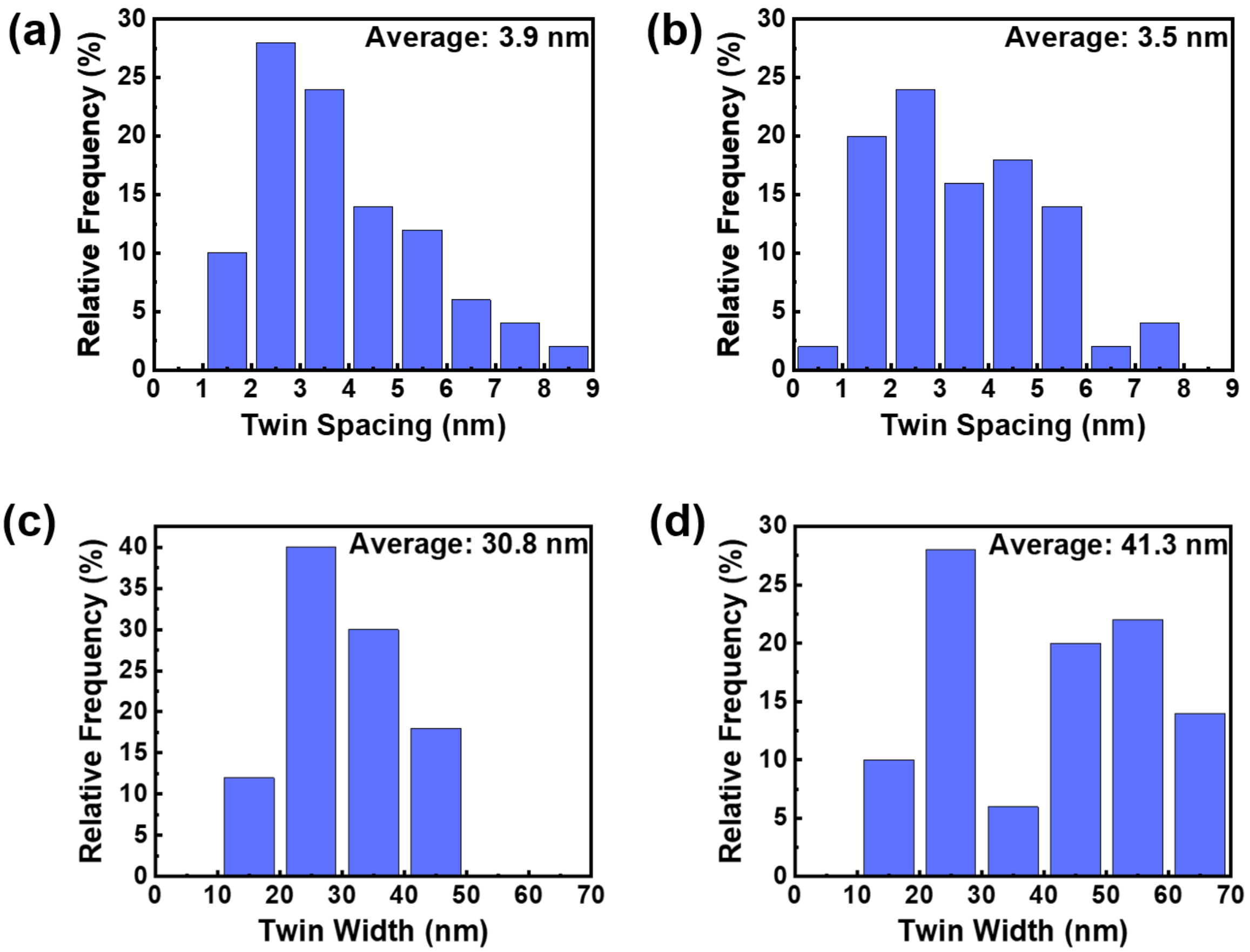

**Figure S5** Distributions of resolved twin spacings and widths measured from representative BF-TEM images. Twin-spacing distributions are shown for (a) Ni-1.5Y and (b) Ni-9.3Cr-1.5Y, while twin-width distributions are shown for (c) Ni-1.5Y and (d) Ni-9.3Cr-1.5Y. Fifty twins were measured for each sample. The mean twin spacings were 3.9 ± 1.7 nm and 3.5 ± 1.7 nm, and the corresponding mean twin widths were 30.8 ± 8.7 nm and 41.3 ± 16.0 nm, respectively.

## Note 6. Additional STEM-EDS characterization of Y grain-boundary segregation

Figure S6 presents additional STEM-EDS measurements from two grain-boundary regions in a nanocrystalline Ni-3.0Y film with an XRD crystallite size of approximately 15 nm. This film was fabricated under different sputtering conditions and is not one of the specific samples reported in the main manuscript. The Y concentration profiles and elemental maps show local Y enrichment at both examined grain boundaries. These measurements support a reproducible tendency for Y grain-boundary segregation across samples with different microstructural and processing conditions, but do not establish universal segregation across all grain-boundary characters or a systematic dependence on the overall Y concentration.

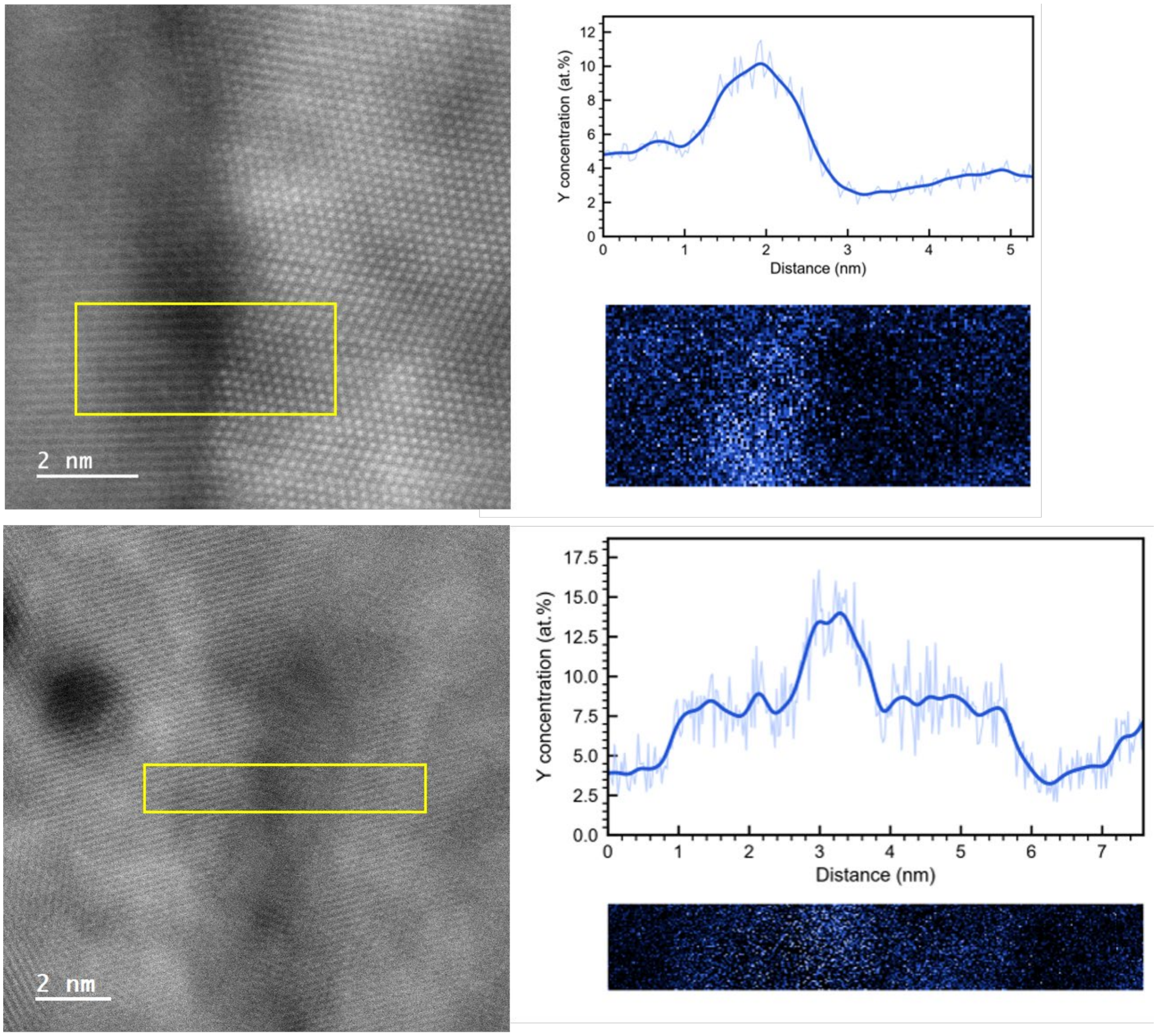

**Figure S6** STEM-EDS characterization of two grain-boundary regions in a nanocrystalline Ni-3.0Y film with an XRD crystallite size of approximately 15 nm. For each dataset, the STEM image is shown on the left, with the yellow rectangle indicating the region used for EDS analysis, while the corresponding Y concentration profile and elemental map are shown on the right. The light-blue lines represent the measured concentration data, and the dark-blue curves show the smoothed profiles. Local Y enrichment is observed at both examined grain boundaries.